\documentclass{LMCS}

\def\dOi{10(4:8)2014}
\lmcsheading%
{\dOi}
{1--40}
{}
{}
{Feb.~27, 2014}
{Dec.~\phantom09, 2014}
{}

\ACMCCS{[{\bf Theory of computation}]: Models of
  computation---Computability---Lambda calculus} 

\usepackage{needspace}
\usepackage{amsmath,amsfonts,amssymb}
\usepackage{proof}
\usepackage{multirow}
\usepackage{url}
\usepackage[matrix,arrow,curve]{xy}
\usepackage[only,llbracket,rrbracket]{stmaryrd}
\usepackage{hyperref}

\newtheorem{theorem}{Theorem}[section]
\newtheorem{lemma}[theorem]{Lemma}
\newtheorem{remark}[theorem]{Remark}
\newtheorem{corollary}[theorem]{Corollary}
\newtheorem{definition}[theorem]{Definition}
\newtheorem{example}[theorem]{Example}
\newtheorem{convention}[theorem]{Convention}
\newenvironment{Corollary}{%
  \Needspace*{3\baselineskip}%
  \corollary
}{\endtheorem}

\newcommand{\recap}[2]{\medskip\noindent{\bf #1 \ref{#2}.}~}
\newcommand{\App}[1]{The details can be found in Appendix~\ref{proof:#1}}

\newcommand{\define}[1]{{\em #1}}
\newcommand{\void}[1]{}

\newcommand{\olin}{\mbox{\sc lineal}}
\newcommand{\oalg}{\mbox{\sc alg}}
\newcommand{\lalin}{\ensuremath{\lambda_{{\it lin}}}}

\newcommand{\xllin}[1]{\ensuremath{\lambda^{\raisebox{0.5ex}{\textrm{\tiny $#1$}}}_{{\it lin}}}}
\newcommand{\xlalg}[1]{\ensuremath{\lambda^{\raisebox{0.5ex}{\textrm{\tiny $#1$}}}_{{\it alg}}}}
\newcommand{\llinred}{\xllin{\rightarrow}}
\newcommand{\lalgred}{\xlalg{\rightarrow}}
\newcommand{\llineq}{\xllin{=}}
\newcommand{\lalgeq}{\xlalg{=}}

\newcommand{\xto}[1]{\ensuremath{\rightarrow_{#1}}}
\newcommand{\simxto}[1]{\ensuremath{\rightarrow_{#1}^{\hspace{-2.2ex}\textrm{\tiny=}}}}
\newcommand{\ssimxto}[1]{\ensuremath{\rightarrow_{#1}^{\hspace{-2.2ex}\textrm{\tiny=}\ *}}}
\newcommand{\fragment}[2]{\ensuremath{{#2}^{\hspace{-1ex}#1}}}
\newcommand{\toolin}{\xto{\mathcal{L}}}
\newcommand{\stoolin}{\ensuremath{\xto{L}^{\ast}}}
\newcommand{\tolinred}{\xto{\ell}}
\newcommand{\toalgred}{\xto{a}}
\newcommand{\tolineq}{\simxto{\ell}}
\newcommand{\toalgeq}{\simxto{a}}

\newcommand{\stoalgred}{\ensuremath{\xto{a}^{\ast}}}
\newcommand{\stolinred}{\ensuremath{\xto{\ell}^{\ast}}}
\newcommand{\symclosure}[1]{\,{#1}^{\leftrightarrow}\,}

\newcommand{\tobv}{\xto{\beta_v}}
\newcommand{\tobn}{\xto{\beta_n}}

\newcommand{\toblinred}{\xto{\ell\cup\beta_v}}

\newcommand{\tobalgred}{\xto{a\cup\beta_n}}
\newcommand{\toblineq}{\simxto{{\ell}\cup\beta_v}}
\newcommand{\tobalgeq}{\simxto{a\cup\beta_n}}

\newcommand{\stoblinred}{\ensuremath{\xto{\ell\cup\beta_v}^{\ast}}}
\newcommand{\stobalgred}{\ensuremath{\xto{a\cup\beta_n}^{\ast}}}
\newcommand{\stoblineq}{\ensuremath{\ssimxto{{\ell}\cup\beta_v}}}
\newcommand{\stobalgeq}{\ensuremath{\ssimxto{a\cup\beta_n}}}

\newcommand{\stolineq}{\ensuremath{\ssimxto{{\ell}}}}
\newcommand{\stoalgeq}{\ensuremath{\ssimxto{a}}}

\usepackage{accsupp} 

\newcommand*{\llbrace}{%
  \BeginAccSupp{method=hex,unicode,ActualText=2983}%
  \textnormal{\usefont{OMS}{lmr}{m}{n}\char102}%
  \mathchoice{\mkern-4.05mu}{\mkern-4.05mu}{\mkern-4.3mu}{\mkern-4.8mu}%
  \textnormal{\usefont{OMS}{lmr}{m}{n}\char106}%
  \EndAccSupp{}%
}
\newcommand*{\rrbrace}{%
  \BeginAccSupp{method=hex,unicode,ActualText=2984}%
  \textnormal{\usefont{OMS}{lmr}{m}{n}\char106}%
  \mathchoice{\mkern-4.05mu}{\mkern-4.05mu}{\mkern-4.3mu}{\mkern-4.8mu}%
  \textnormal{\usefont{OMS}{lmr}{m}{n}\char103}%
  \EndAccSupp{}%
}

\newcommand{\wt}[1]{\llbracket{#1}\rrbracket}
\newcommand{\cps}[1]{\llbrace{#1}\rrbrace}
\newcommand{\pair}[2]{\left\langle #1,#2\right\rangle}

\newcommand{\cont}[1]{\lambda f\,\left(f\right)~#1}
\newcommand{\uncont}[2]{\lambda #1\,({#2}\,#1)}
\newcommand{\tapp}[5]{\lambda{#1}\,({#4})~\lambda{#2}\,({#5})~\lambda{#3}\,(({#2})~{#3})~{#1}}
\newcommand{\tappfgh}{\tapp{f}{g}{h}}
\newcommand{\tappcps}[2]{\lambda{f}\,({#1})~\lambda{g}\,((g)~{{#2}})~f}

\newenvironment{absolutelynopagebreak}
  {\par\nobreak\vfil\penalty0\vfilneg
   \vtop\bgroup}
  {\par\xdef\tpd{\the\prevdepth}\egroup
   \prevdepth=\tpd}

\makeatletter 
\def\mynobreakpar{\par\nobreak\@afterheading} 
\makeatother

\allowdisplaybreaks

\begin{document}

\title[CbV, CbN and the vectorial behaviour of the algebraic $\lambda$-calculus]{Call-by-value, call-by-name and the vectorial behaviour of the algebraic $\lambda$-calculus}

\author[A.~Assaf]{Ali Assaf\rsuper a}
\address{{\lsuper{a}}\'Ecole Polytechnique\\
Route de Saclay\\
91120 Palaiseau\\
France}
\address{{\lsuper{a}}INRIA\\
23 avenue d'Italie\\
CS 81321\\
75214 Paris Cedex 13\\
France}
\email{ali.assaf@inria.fr}

\author[A.~D\'iaz-Caro]{Alejandro D\'iaz-Caro\rsuper b}
\address{{\lsuper b}Universidad Nacional de Quilmes\\
Roque S\'aenz Pe\~na 352\\
1876 Bernal, Buenos Aires\\
Argentina}
\email{alejandro@diaz-caro.info}
\thanks{\lsuper{b} Partially supported by the French-Argentinian Laboratory in Computer Science INFINIS}

\author[S.~Perdrix]{Simon Perdrix\rsuper c}
\address{{\lsuper c}CNRS \& LORIA\\
615, rue du Jardin Botanique\\
BP-101\\
54602 Villers-l\`es-Nancy\\
France}
\email{simon.perdrix@loria.fr}

\author[C.~Tasson]{Christine Tasson\rsuper d}
\address{{\lsuper{d,e}}PPS, Universit\'e Paris-Diderot -- Paris 7\\
CNRS UMR 7126\\
75205 Paris Cedex 13\\
France}
\email{christine.tasson@pps.univ-paris-diderot.fr, benoit.valiron@monoidal.net}

\author[B.~Valiron]{Beno\^it Valiron\rsuper e}
\address{{\lsuper{e}} I2M, Universit\'{e} Aix-Marseille,\\
CNRS UMR 7373, 
Campus de Luminy, case 907, F–13288 Marseille\\
France}
\thanks{{\lsuper{e}} Partially supported by the ANR project ANR-2010-BLAN-021301 LOGOI.}


\begin{abstract} 
  We examine the relationship between the {\em algebraic
  $\lambda$-calculus}, a fragment of the differential
  $\lambda$-calculus and the \emph{linear-algebraic
  $\lambda$-calculus}, a candidate $\lambda$-calculus for quantum
  computation. Both calculi are algebraic: each one is equipped with
  an additive and a scalar-multiplicative structure, and their set of
  terms is closed under linear combinations. However, the two
  languages were built using different approaches: the former is a
  call-by-name language whereas the latter is call-by-value; the
  former considers algebraic equalities whereas the latter approaches
  them through rewrite rules.

  In this paper, we analyse how these different approaches relate to one
  another. To this end, we propose four canonical languages based
  on each of the possible choices: call-by-name versus call-by-value,
  algebraic equality versus algebraic rewriting. We show that the
  various languages simulate one another.
  Due to subtle interaction between beta-reduction and algebraic
  rewriting, to make the languages consistent some additional
  hypotheses such as confluence or normalisation might be required. We carefully devise
  the required properties for each proof, making them general enough
  to be valid for any sub-language satisfying the corresponding
  properties. 
\end{abstract}

\maketitle

\section{Introduction}

Two algebraic versions of the $\lambda$-calculus arise independently in distinct contexts: the algebraic $\lambda$-calculus ($\oalg$)~\cite{VauxMSCS09} and the linear algebraic $\lambda$-calculus ($\olin$)~\cite{ArrighiDowekRTA08}. 
Both languages are extensions of $\lambda$-calculus where linear combinations of terms are also terms.
The former has been introduced in the context of linear logic as a fragment of the differential $\lambda$-calculus~\cite{EhrhardRegnierTCS03}: the algebraic structure allows to gather in a non deterministic manner different terms, {\em i.e.}\ each term in the linear combination represents one possible execution. The latter has been introduced as a candidate $\lambda$-calculus for quantum computation: in $\olin$, a linear combination of terms reflects the phenomenon of superposition, {\em i.e.}\ the capability for a quantum system to be in two or more states at the same time.
Our purpose is to study the connections between the two systems.

In both languages, functions which are linear combinations of terms are interpreted pointwise: $(\alpha. f + \beta. g)~x = \alpha.(f)~x+\beta.(g)~x$, where ``$.$'' denotes the scalar multiplication. The two languages differ in the treatment of the arguments. In $\olin$, 
in order to deal with the algebraic structure, any function is considered as a linear map: $(f)~(\alpha.x + \beta.y) \to^* \alpha.(f)~x+\beta.(f)~y$, reflecting the fact that any quantum evolution is a linear map. It reflects a call-by-value behaviour in the sense that the argument is evaluated until one has a base term.
Conversely, $\oalg$ has a call-by-name evolution: $(\lambda x\, M)~N \to M[x:= N]$, without any restriction on $N$. As a consequence, the evolutions are different as illustrated by the following example. In $\olin$, $(\lambda x\,(x)~x)~(\alpha.y+\beta.z) \to^* \alpha.(y)~y+ \beta.(z)~z$ while in $\oalg$, $(\lambda x\,(x)~x)~(\alpha.y+\beta.z) \to (\alpha.y+\beta.z)~(\alpha.y+\beta.z)= \alpha^2.(y)~y+ \alpha\beta.(y)~z +\beta\alpha.(z)~y+ \beta^2.(z)~z$. 

Because they were designed for different purposes, another difference
appears between the two languages: the way the algebraic part of the
calculus is treated. In $\olin$, the algebraic structure is captured
with a rewrite system, whereas in $\oalg$ terms are considered up to
algebraic equivalence.

The two choices -- call-by-value versus call-by-name and algebraic
equality versus algebraic reduction -- allow one to construct four
possible calculi. We name them \llinred, \llineq, \lalgred, and \lalgeq. See Figure~\ref{tab:linalg} where they
are presented according to their evaluation policy and the
way they take care of the algebraic part of the language.
\begin{figure}[b]
  \def\arraystretch{1.5}
  \begin{tabular}{|c|c|c|}
    \hline
    & call-by-name  &  call-by-value \\
    \hline
    algebraic & \multirow{2}{*}{\lalgred} & \multirow{2}{*}{\llinred}\\[-0.2cm]
    reduction 
    & & \\
    \hline
    algebraic & \multirow{2}{*}{\lalgeq} & \multirow{2}{*}{\llineq}\\[-0.2cm]
    equality & & \\
    \hline
  \end{tabular}\\[1.5ex]
  \caption{The four algebraic $\lambda$-calculi.}
  \label{tab:linalg}
\end{figure}

Inspired by $\olin$ and $\oalg$, the operational semantics of these four languages differ slightly from the original ones to better emphasise their characteristics:
the reduction strategy and the handling of algebraic structure.

A first modification is that in all four languages,  we avoid reduction under lambda abstractions.
As a consequence, contrary to $\oalg$, the $\lambda$-abstraction is not linear anymore: $\lambda x\,(\alpha.M +\beta. N) \neq \alpha. \lambda x\, M + \beta. \lambda x\,N$.  This restriction is a common restriction: reducing under $\lambda$ could be considered as ``optimising the program''.

Concerning $\llinred$ and $\llineq$, 
restrictions originally imposed in $\olin$ on the rewrite system to ensure confluence are replaced by restrictions making $\llinred$ and $\llineq$ call-by-value also in the algebraic part. 
For example, in the rule
$(M+N)~L\to (M)~L+(N)~L$ the condition that $M+N$ be closed-normal is replaced
by the restriction of $L$ to values. Notice that even in the original language $\olin$, waiving the restrictions makes sense when confluence can be ensured by other means, see {\em e.g.}~\cite{ArrighiDiazcaroQPL09,ValironDCM10}.
Since this change in the strategy is not trivial, we prove that
$\olin$ and $\llinred$ share the same behaviour, result formalized in Theorems~\ref{thm:LLINtoLIN} and \ref{thm:LINtoLLIN}.

Our contribution in this paper is first to introduce the four canonical languages related to the original languages $\olin$ and $\oalg$, and then to show the relation between $\olin$ and $\oalg$, by proving the simulation of all these languages.
We show that call-by-value algebraic $\lambda$-calculi simulate
call-by-name ones and {\em vice versa} by extending the continuation
passing style (CPS) translation~\cite{PlotkinTCS75} to the algebraic case. We
also provide simulations between algebraic equality and algebraic
reduction in both directions. The simulations we prove are summed up
in Figure~\ref{fig:relation}. The solid arrows stand for theorems that
do not require confluence or normalisation in their hypothesis whereas the dotted
arrows stand for theorems requiring confluence, and the dashed arrows for theorems requiring strong normalisation. The star in $\oalg$ and $\olin$ means that the languages do not allow reduction under $\lambda$.

\subsection*{Related works}
This paper connects two collections of works: one stemming from linear
logic, the other one from quantum computation.

An important line of work in linear logic is concerned with the
development of quantitative semantics: semantics where types are
interpreted as (topological) vector spaces and lambda-terms as power
series. Such semantics include for example finiteness
spaces~\cite{EhrhardMSCS05,TassonTLCA09}, and K\"{o}the
spaces~\cite{EhrhardMSCS03}. The models naturally question the
introduction of new constructs into the logic, to account for the
structure stemming from the vector spaces. In a seminal paper, Ehrhard
and Regnier~\cite{EhrhardRegnierTCS03} develop a lambda-calculus with
a structure of module and a differential operator capturing the inner
structure of finiteness spaces. Because of the original understanding
of lambda-terms as power-series, the resulting lambda-calculus ends up
naturally call-by-name: $(\lambda x\,(f\,x)\,x)(y+z)$ is
$(f\,(y+z))(y+z)$ and not $(f\,y)\,y + (f\,z)\,z$, and lambda-terms
are considered modulo the equations of a module. Its properties were later
extensively studied by Vaux~\cite{VauxRTA07,VauxMSCS09}.

These results also shed some light on another kind of
lambda-calculus: the resource calculus~\cite{BoudolINRIA93}. In the
resource calculus 
the term $(\lambda x\,(f\,x)\,x)[y,z]$ reduces to the non-deterministic superposition $(f\,y)\,z+(f\,z)\,y$.
As a non-deterministic calculus, the resource
calculus is equipped with a sum (the non-deterministic choice), and
many of the tools and techniques developed along with quantitative
semantics are applicable. For example, resource lambda-terms can be
equipped with a finiteness structure~\cite{EhrhardLICS10}, and can be
made in relation with the representation as power series of
lambda-terms~\cite{EhrhardLMCS12,BoudesCSL13}.

\smallskip
Another line of work focuses on lambda-calculi equipped with a module
structure: works stemming from quantum computation. In quantum
computation, the idea of working with linear combinations of terms is
a natural one. Indeed, the state of a quantum register is modelled as
a linear (complex) combination of classical states. If the states in
superposition represent terms of a lambda-calculus, one naturally
ends up with linear combinations of lambda-terms. However, it turns
out not to be so simple: because quantum operations are supposed to
preserve norm and orthogonality, van Tonder~\cite{vanTonderSIAM04}
shows that a general lambda-calculus encoded on quantum states with a
reduction defined as a unitary map (i.e. an internal operation on
quantum states), no non-trivial combinations of terms can occur,
essentially falling back on a classical lambda-calculus manipulating
quantum data.

In order to eventually figure out a solution to this problem, an
obvious first-step is to study what would happen in a more general
setting, and first study the computational aspect of vector spaces and
lambda-calculi with vectorial structures. Arrighi and Dowek first
proposed a computational definition of vector
space~\cite{ArrighiDowekWRLA04} where the equations of vector spaces
are oriented: a vector becomes a term evolving according to a
confluent rewrite system. Later, they propose~\cite{ArrighiDowekRTA08}
one of the lambda-calculus with a vectorial structure on which we
shall concentrate on this paper: $\olin$. The philosophy behind the
presentation of the language
is the following. First, beside associativity and commutativity,
all the rewrite rules are oriented, whether they come from the regular
structure of lambda-calculus or from the vector space structure. Then,
the language is call-by-value in spirit: being an abstraction over
quantum computation, a lambda-term is both an operator and a
state. Therefore, unlike in the linear-logic approach, $(\lambda
x\,(f\,x)\,x)(y+z)$ indeed corresponds to $(f\,y)\,y + (f\,z)\,z$.

Along with the description of the language, Arrighi and Dowek expose a
main issue: while the language na\"{i}vely enjoy local confluence, the
equational theory is inconsistent. They propose a solution by
constraining the rewrite system, while other pieces of work explore other
options.
The first option proposed is by using a type system enforcing strong
normalization, and thus recovering consistency. Several type systems
capturing the algebraic aspect of the language have been developed to
this end: a system-F type system with
scalars~\cite{ArrighiDiazcaroQPL09,ArrighiDiazcaroLMCS12}, a system-F
with a
sum-type~\cite{BuirasDiazcaroJaskelioffLSFA11,DiazcaroPetitWoLLIC12},
a so-called ``vectorial'' type system, where linear combinations of
types are valid
types~\cite{ArrighiDiazcaroValironDCM11,ArrighiDiazcaroValiron13}. The
second option proposed consists in minimally modifying the rewrite
system and adding a type system; consistency is obtained through
the development of a model~\cite{ValironDCM10,ValironMSCS13}.

All these approaches are stepping stones working towards a unique goal:
better understand the connection between lambda-calculus and vectorial
structures,
to eventually come back with an approach different from the one of van
Tonder~\cite{vanTonderSIAM04} for how to reconciliate
lambda-calculi with vectorial structures and quantum computation.
In~\cite{ValironQPL10}, one of the author hints at how one could do
this, by using an intermediate language that can both be ``run'' by a
quantum circuit and interpreted as a vectorial lambda-term.

\smallskip 
A preliminary work-in-progress version of this paper has been presented in workshops in two parts in \cite{DiazcaroPerdrixTassonValironHOR10} (informal) and \cite{AssafPerdrixDCM11}.
The former briefly sketches a
preliminary connection between call-by-value and call-by-name approaches in
lambda-calculi with vectorial structures, using thunks
to simulate call-by-name with call-by-value. In the current paper, we use
instead CPS as it offers a nicer symmetric
connection. The technique for the completeness of the simulation from
call-by-value to call-by-name was developed in~\cite{AssafPerdrixDCM11}.
In these preliminary presentations, there are some small differences in the reduction rules with respect to the current work. The reason is that in the current work we tried to present the most general reduction strategy, adding as less conditions as possible. For example, in \cite{DiazcaroPerdrixTassonValironHOR10} the left linearity rule is $(U+V)\,W\to U\,W+V\,W$
and requires all the subterms to be values in order to reduce, while in the
present paper it is $(M+N)\,V \to M\,V+N\,V$ and only requires the argument to be a value.

\begin{figure}
  \centering
  $\xymatrix@C=4ex@R=4ex{
    &&&&&
    \olin^*
    \ar@{-->}@/_4ex/[ld]|(0.6){\textrm{Th.\,\ref{thm:LINtoLLIN}}}
    \\
    &
    \lalgred
    \ar@{->}@/_4ex/[ddd]|(0.6){\textrm{Th.\,\ref{th:redeq}}}
    \ar@{->}@/_4ex/[rrr]|{\textrm{Th.\,\ref{th:sim2}}}
    &
    &
    & 
    \llinred
    \ar@{-->}@/_4ex/[ru]|(0.6){\textrm{Th.\,\ref{thm:LLINtoLIN}}}
    \ar@{->}@/_4ex/[lll]|{\textrm{Th.\,\ref{th:sim}}}
    \ar@{->}@/_4ex/[ddd]|(0.6){\textrm{Th.\,\ref{th:redeq2}}} 
    &
    \\&&&&
    \\&&&&\\
    &
    \lalgeq 
    \ar@{->}@/_4ex/[rrr]|{\textrm{Th.\,\ref{th:sim2eq}}}
    \ar@{.>}@/_4ex/[uuu]|(0.6){\textrm{Th.\,\ref{thm:eqred2}}}
    &
    &
    &
    \llineq
    \ar@{->}@/_4ex/[lll]|{\textrm{Th.\,\ref{th:simeq}}}
    \ar@{.>}@/_4ex/[uuu]|(0.6){\textrm{Th.\,\ref{thm:eqred}}}
    &
    \\
    \oalg^*\ar@{=}[ru]
    &&&&&
  }$\\[1.5ex]
  \small
  $A\to B$ means ``$A$ is simulated by $B$''
  \caption{Relations between the languages. }
  \label{fig:relation}
\end{figure}

\subsection*{Plan of the paper.}
In Section~\ref{sec:originals}, we present the original calculi $\olin$ and $\oalg$.
In Section~\ref{sec:alglam}, we define the set of terms and the rewrite systems we consider in the paper.
In Section~\ref{sec:relation-orig}, we establish the relation between the original setting and the setting used in this paper.
In Section~\ref{sec:consistency}, we discuss the confluence 
of the algebraic rewrite systems. 
Section~\ref{sec:sim} is concerned with the actual simulations. In Section~\ref{sec:redeq} we consider the correspondence between algebraic reduction and algebraic equality whereas in Sections~\ref{subsec:lintoalg} to~\ref{subsec:algtolin-completeness} we consider the distinction call-by-name versus call-by-value. In Section~\ref{subsec:compose}, we show how the simulations can compose to obtain the correspondence between any two of the four languages. 
In Section~\ref{sec:concl} we conclude by providing some paths for future work.
Omitted and sketched proofs are fully developed in the appendix.

\section{The languages}
In this section we present all the languages: the original setting, and our standardised versions of the algebraic calculi.
\subsection{The original setting}\label{sec:originals}
The language \olin\ was first presented in~\cite{ArrighiDowekRTA08} as
summarised in Figure~\ref{fig:olin}. The rewrite system is defined by
structual induction on the left-hand-side. The factorisation and
application rules ask for a particular subterm of the left-hand-side
to {\em not} reduce (conditions (*) and (**)). Because of the
inductive definition, this is indeed well-defined.
These conditions (*) and (**) in particular ensure confluence: we
refer the reader to the original paper~\cite{ArrighiDowekRTA08} for
more details.

\begin{figure}
  {
    \hrule\vspace{1pt}\hrule
    $$\begin{array}[t]{l@{\hspace{1.5cm}}r@{\ ::=\quad}l}
      \textrm{\itshape Terms:} & M,N,L & B\ |\ (M)~N\ |\ 0\ |\ \alpha.M\ |\ M+N\\
      \textrm{\itshape Base terms:} & B & x\ |\ \lambda x\,M
    \end{array}$$

    The rewrite system is defined inductively on the size of the left-hand-side.\\[2ex]
    
    \begin{tabular}{p{5.6cm}p{5.6cm}} 
      \noindent \emph{Elementary rules:}

      \noindent $M+0\to M$,

      \noindent $0.M\to0$,

      \noindent $1.M\to M$,

      \noindent $\alpha.0\to0$,

      \noindent $\alpha.(M+N)\to\alpha.M+\alpha.N$.
      &
      \noindent \emph{Beta reduction:}

      \noindent $(\lambda x\,M)~B\to M[x:=B]$.
      \medskip
      
      \noindent \emph{Factorisation rules},
      
      \noindent $\alpha.M+\beta.M\to(\alpha+\beta).M$  (*),
      
      \noindent $\alpha.M+M\to(\alpha+1).M$  (*),
      
      \noindent $M+M\to(1+1).M$  (*),
      
      \noindent $\alpha.(\beta.M)\to(\alpha\beta).M$.
    \end{tabular}
    \medskip

    \begin{tabular}{p{6cm}p{5.6cm}}
      \multicolumn{2}{l}{\emph{Application rules:}} \\
      \noindent $(M+N)~L\to (M)~L+(N)~L$ (**),

      \noindent $(L)~(M+N)\to (L)~M+(L)~N$ (**),

      \noindent $(\alpha.M)~N\to\alpha.(M)~N$ (*), &

      \noindent $(N)~(\alpha.M)\to\alpha.(N)~M$ (*),

      \noindent $(0)~M\to0$,

      \noindent $(M) 0\to0$.
    \end{tabular}
  }
  \bigskip

  \noindent where $+$ is an associative-commutative ({\em AC}) symbol,
  $\alpha, \beta\in\mathcal{S}$, with $(\mathcal{S},+,\times)$ a
  commutative ring and (see Section~\ref{sec:originals} for details)\\
  (*) these rules apply only if $M$ is closed and does {\em not} reduce.\\
  (**) these rules apply only if $M+N$ is closed and does {\em not} reduce.
  \medskip

  \hrule\vspace{1pt}\hrule
  \caption{Syntax and reduction rules of \olin\ as they first appeared in~\cite{ArrighiDowekRTA08}}
  \label{fig:olin}
\end{figure}

The language \oalg\ was first presented in~\cite{VauxMSCS09} as
summarised in Figure~\ref{fig:oalg}. Notice the equality instead of
rewrite in the treatment of the algebraic part of the language. Also
note that it does not ask for any particular side-condition as it is
the case in $\olin$.
\begin{figure}
  {
    \hrule\vspace{1pt}\hrule
    $$\begin{array}[t]{l@{\hspace{1.5cm}}r@{\ ::=\quad}l}
      \textrm{\itshape Terms:} & M,N,L & x\ |\ \lambda x\,M\ |\ (M)~N\ |\ 0\ |\ \alpha.M\ |\ M+N\\
    \end{array}$$
    \begin{tabular}{p{5.6cm}p{5.6cm}} 
      \noindent \emph{Axioms of commutative monoid:}

      \noindent $M+0 = M$,

      \noindent $M+N = N+M$,

      \noindent $(M+N)+L = M+(N+L)$
      &
      \noindent \emph{Beta reduction:}

      \noindent $(\lambda x\,M)~N\to M[x:=N]$.
      \\
      &\\
      \noindent \emph{Axioms of module over the ring:}

      \noindent $\alpha.(M+N) = \alpha.M+\alpha.N$,

      \noindent $\alpha.M+\beta.M = (\alpha+\beta).M$,

      \noindent $\alpha.(\beta.M) = (\alpha\beta).M$,

      \noindent $1.M = M$,

      \noindent $0.M = M$,

      \noindent $\alpha.0 = 0$.
      &
      \emph{Linearity in the $\lambda$-calculus:}

      \noindent $\lambda x\,0 = 0$,

      \noindent $\lambda x\,(\alpha.M) = \alpha.\lambda x\,M$,

      \noindent $\lambda x\,(M+N) = \lambda x\,M+\lambda x\,N$,

      \noindent $(0)~M = 0$,

      \noindent $(\alpha.M)~N = \alpha.(M)~N$,

      \noindent $(M+N)~L = (M)~L+(N)~L$.
    \end{tabular}
  }
  \medskip

  \hrule\vspace{1pt}\hrule
  \caption{Syntax and reduction rules of \oalg\ as they first appeared in~\cite{VauxMSCS09}.}
  \label{fig:oalg}
\end{figure}

Besides the treatment of the rewrite rules, we notice
that \olin\ has a call-by-value behaviour while \oalg\ has a
call-by-name behaviour. However, we believe that it is not possible to
draw in a straightforward manner a direct CPS translation between
\olin\ and \oalg. The intuitions behind this claim is that Vaux's
calculus uses algebraic equalities in an unrestricted manner, whereas
Lineal uses oriented algebraic rewrite rules in a very constraints
manner (because of the side-conditions (*) and (**) on rules in
Figure~\ref{fig:olin}). A CPS translation from $\oalg$ to $\olin$
would therefore have to force algebraic reductions sending for example
$\alpha.x+\beta.x$ to $(\alpha+\beta).x$, which is feasable in
$\oalg$, but not in $\olin$. The translation would also have to deal
with the case that $\oalg$ have unoriented algebraic rewrites, whereas
$\olin$ has oriented ones. So the CPS translation would have to be
able to map the translation of $0$ to the translation of $0.M$, for
all $M$.
 
In this paper, we fully desribe the connection between these two
languages using a CPS translation, but the path we follow is therefore
to first standardise the languages before establishing the
simulations.

\subsection{Standardised algebraic $\lambda$-calculi}\label{sec:alglam}

The original languages $\olin$ and $\oalg$ made particular assumptions
both on the reduction strategy and the handling of algebraic structure
under the reduction.
In this paper, we consider separately the distinction
call-by-name/call-by-value and the distinction algebraic
equality/algebraic reduction. We
develop therefore four languages: a call-by-value language $\llineq$
with algebraic equality, a call-by-value language $\llinred$ with
algebraic reduction, a call-by-name language $\lalgeq$ with algebraic
equality and a call-by-name language $\lalgred$ with algebraic
reduction. 
These four languages are summarised in Figure~\ref{tab:linalg}.

The languages share the same syntax, defined as follows:
\[\begin{array}{rcll}
    M,N,L&::=& V ~|~ (M)~N ~|~ \alpha.M ~|~M+N&\textrm{(terms),}\\
    U,V,W&::=& 0~|~B~|~\alpha.V~|~V+W&\textrm{(values),}\\
    B&::=& x~|~ \lambda x\,M&\textrm{(basis terms),}
\end{array}\]
where $\alpha$ ranges over a ring, the \define{ring of scalars}.
We use the notation $M-N$ as a shorthand for
$M+(-1).N$.
Note that we could have asked for a semiring instead; in
fact we shall see in Section~\ref{sec:consistency} that the analysis we
develop here can be adapted to semirings of scalars.

We summarise in Figure~\ref{tab:RW} all the
rewrite rules.
The rules are grouped with respect to their intuitive meaning. 
We use the usual notation regarding rewrite systems: Given a rewrite system $R$, we write $R^*$ for its reflexive and transitive closure. That is, $xR^*y$ is valid if $y=x$ or if there exists a rewrite sequence $x\,R\,x_1\,R\,\cdots \,R\,x_n\,R\,y$ linking $x$ and $y$.
We write $\symclosure R$ for the symmetric closure of $R$, that is, the relation that satisfies $x\symclosure{R}y$ if and only if $x\,R\,y$ or $y\,R\,x$.

\begin{figure}
  \begin{center}
    {
      $\begin{array}{rcl@{\qquad}rcl}
	\hline
	\hline
	\multicolumn{6}{c}{\rule{0ex}{4ex}{\textrm{\sc Specific rules for $\lalgred$ and $\lalgeq$}}}\\
	\hline\\[-1ex]
	\multicolumn{3}{c}{\textrm{Call-by-name ($\beta_n$)}}
	&
	\multicolumn{3}{c}{\textrm{\sc Linearity of the application ($A$)}}\\[1.5ex]
	(\lambda x\,M)~N &\to& M[x:=N] 
	& 
	(M+N)~L &\to& (M)~L + (N)~L \\
	&&&
	(\alpha.M)~N &\to& \alpha.(M)~N\\
	&&&
	(0)~M &\to& 0\\[1.5ex]
	\hline
	\hline
	\multicolumn{6}{c}{\rule{0ex}{4ex}{\textrm{\sc Specific rules for $\llinred$ and $\llineq$}}}\\
	\hline\\[-1ex]
	\multicolumn{3}{c}{\textrm{Call-by-value ($\beta_v$)}}
	&
	\multicolumn{3}{c}{\textrm{\sc Context rule ($\xi_{\lalin}$)}}\\[1.5ex]
	(\lambda x\,M)~B &\to& M[x:=B] 
	& 		
	\multicolumn{3}{c}{\infer{(V)~M\to (V)~M'}{M\to M'}}\\
	\multicolumn{6}{c}{\textrm{\sc Linearity of the application}}\\[1.5ex]
	\multicolumn{3}{c}{\textrm{Left linearity ($A_l$)}}
	&
	\multicolumn{3}{c}{\textrm{Right linearity ($A_r$)}}\\[1.5ex]
	(M+N)~V &\to& (M)~V + (N)~V 
	&
	(B)~(M+N) &\to& (B)~M + (B)~N\\
	(\alpha.M)~V &\to& \alpha.(M)~V
	&
	(B)~(\alpha.M) &\to& \alpha.(B)~M\\
	(0)~V &\to& 0 
	&
	(B)~0 &\to& 0\\[1.5ex]
	\hline
	\hline
	\multicolumn{6}{c}{\rule{0ex}{4ex}{\textrm{\sc Common rules}}}\\
	\hline\\[-1ex]
	\multicolumn{6}{c}{\textrm{\sc Ring rules ($L={\it Asso}\cup{\it Com}\cup F\cup S$)}}\\[1.5ex]
	\multicolumn{3}{c}{\textrm{Associativity (${\it Asso}$)}}
	&
	\multicolumn{3}{c}{\textrm{Commutativity (${\it Com}$)}}\\[1.5ex]
	M+(N+L) &\to& (M+N)+L
	&
	M+N &\to& N+M\\
	(M+N)+L &\to& M+(N+L)
	&
	\\[1.5ex]
	\multicolumn{3}{c}{\textrm{Factorisation ($F$)}}
	&
	\multicolumn{3}{c}{\textrm{Simplification ($S$)}}\\[1.5ex]
	\alpha.M + \beta.M &\to& (\alpha+\beta).M
	&
	\alpha.(M+N) &\to& \alpha.M + \alpha.N\\
	\alpha.M + M &\to& (\alpha + 1).M
	&
	1.M &\to& M\\
	M + M &\to& (1 + 1).M
	& 
	0.M &\to& 0\\
	\alpha.(\beta.M) &\to& (\alpha\beta).M
	&
	\alpha.0 &\to& 0\\
	&&
	&
	0 + M &\to& M\\[1ex]
	\multicolumn{6}{c}{\textrm{\sc Context rules ($\xi$)}}\\[1.5ex]
	\multicolumn{2}{c}{\infer{(M)~N\to (M')~N}{M\to M'}}
	&
	\infer{M + N\to M' + N}{M\to M'}
	&
	\multicolumn{2}{c}{\infer{M + N\to M + N'}{N\to N'}}
	&
	\infer{\alpha.M\to \alpha.M'}{M\to M'}\\[1.5ex]
	\\
	\hline\hline
      \end{array}$
    }
  \end{center}
  \caption{Rewrite rules with $U,V$ and $W$ values, $B$ a basis term,
  $M,M',N,N'$ and $L$ any terms.}
  \label{tab:RW}
\end{figure}

\begin{definition}\label{def:lang} We use the following notations for
  the rewrite systems obtained by combining the rules described in
  Figure \ref{tab:RW}:
  \[\begin{array}{rcl@{\hspace{0.75cm}}rcl@{\hspace{0.75cm}}rcl}
      \toalgred& := &A \cup L \cup \xi &\tolinred&:=&A_l \cup A_r \cup L \cup \xi \cup \xi_{\lalin} &\tobv&:=&\beta_v \cup \xi \cup \xi_{\lalin}\\
      \toalgeq&:=&\symclosure{(\toalgred)}&\tolineq&:=&\symclosure{(\tolinred)}&\tobn&:=&\beta_n \cup \xi
  \end{array}\]
  We hence define the following four languages and their associated
  rewrite systems:

  \medskip
  \begin{center}
    \begin{tabular}{|c|c|}\hline
      Language & Corresponding Rewrite System\\
      \hline
      $\llinred$ & $\toblinred:=(\tolinred)\cup(\tobv)$\\
      $\llineq$ & $\toblineq:=(\tolineq)\cup(\tobv)$\\
      $\lalgred$ & $\tobalgred := (\toalgred)\cup(\tobn)$\\
      $\lalgeq$ &$\tobalgeq:=(\toalgeq)\cup(\tobn)$\\\hline
    \end{tabular}
  \end{center}
  The ``$=$'' symbol over the arrow is to refer to $\llineq$ and $\lalgeq$, it does not mean that the beta rule is also reflexive.

  \medskip\noindent For each language, the notion of {\em normal form}
  is defined as follows: 
  \begin{itemize}
    \item a term $M$ is in {\em normal form with respect to $\toblinred$}
      (respectively $\tobalgred$) if there is no term $M$ such that
      $M\toblinred N$ (respectively $M\tobalgred N$).
    \item a term $M$ is in {\em normal form with respect to $\toblineq$}
      (respectively $\tobalgeq$) if for all terms $M$ such that
      $M\toblineq N$ (respectively $M\tobalgeq N$), $N\tolineq M$ (respectively $N\toalgeq M$).
  \end{itemize}
  In other word, a term is in normal form is it does not reduce,
  potentially modulo the algebraic equalities inscribed within the
  rewrite system.
\end{definition}

\subsection{Relation between the standardised languages and the original settings}\label{sec:relation-orig}
Let $\olin^*$ and $\oalg^*$ be $\olin$ and $\oalg$ respectively without reduction under $\lambda$. Not reducing under $\lambda$ in $\oalg$ implies also removing the first three rules of ``Linearity in the $\lambda$-calculus''.
Also, since we consider $\lalgred$ with algebraic rewrite rules instead of the equalities used in $\oalg$, we need two extra rules: $\alpha.M + M \to (\alpha+1).M$ and $M+M \to (1+1).M$. These rules were not needed with equalities, because $M = 1.M$.
It is easy to check that $\oalg^*$ and $\lalgeq$ are actually the same language. However, $\llinred$ differs from $\olin^*$. Besides eliminating conditions (*) and (**) ({\em cf.}~Figure~\ref{fig:olin}), we enforce a full call-by-value strategy in the algebraic part. The following results (Theorems~\ref{thm:LLINtoLIN} and \ref{thm:LINtoLLIN}) show that $\llinred$ and $\olin^*$ still have the same behaviour, in the sense that if $M$ reduces to $N$ either in $\olin^*$ or in $\llinred$, then there exists $L$ such that both $M$ and $N$ reduce to $L$ in the other language. We use the notation $\toolin$ for reductions in $\olin^*$.

Terms in normal form in $\olin^*$ are simply non-reducing terms,
whereas normal forms in $\oalg^*$ are non-reducing terms {\em modulo}
the equalities on Figure~\ref{fig:oalg}.

\begin{theorem}
  \label{thm:LLINtoLIN}
  If $M$ is closed and strongly normalising in $\olin^*$ and $M\toblinred N$,
  then there exists $L$ such that $M\toolin^* L$ and $N\toolin^* L$.
\end{theorem}
\begin{proof}
  The proof is done by induction on the $\toblinred$ rewrite relation. \App{LLINtoLIN}.
\end{proof}

\begin{theorem}
  \label{thm:LINtoLLIN}
  If $M$ is closed and strongly normalising in $\olin$ and  $M\toolin N$, then there exists $L$ such that $M\stoblinred L$ and $N\stoblinred L$.
\end{theorem}
\begin{proof}
  We only need to verify the seven differing rules between the two languages.
  Notice that the normal form of a closed term is a value $V$.
  The full details can be found in Appendix~\ref{proof:LINtoLLIN}.
\end{proof}

\subsection{Discussion on consistency and confluence}\label{sec:consistency}

It is known that, without restrictions, both algebraic reductions and
algebraic equalities cause problems of consistency, albeit
differently. The original languages solve these problems using
different techniques.

Let $Y_M=(\lambda x\,(M+(x)~x))~\lambda x\,(M+(x)~x)$. In a system
with algebraic reduction such as $\olin$, $\llinred$, or $\lalgred$, the term
$Y_M-Y_M$ reduces to $0$, but also reduces to $M+Y_M-Y_M$ and hence to
$M$, breaking confluence.  In a system with algebraic equalities such
as $\oalg$, $\llineq$, or $\lalgeq$, it
is even worse: for any terms $M$ and $N$, because algebraic
rewrites are not oriented, the following reductions hold
\[
  M \toalgeq M+0
  \toalgeq  M+(Y_{N-M}-Y_{N-M})
  \tobalgeq  M+((N-M+Y_{N-M})-Y_{N-M})
  \stoalgeq  N
\]
making any term $M$ reduce to any other term $N$.

To solve this issue, several distinct techniques can be used to make
an algebraic calculus confluent. In the original presentation of
$\olin$ (see Figure~\ref{fig:olin}), restrictions on reduction rules
are introduced.  For example, $\alpha.M+\beta.M\to (\alpha+\beta).M$
only if $M$ is closed and normal (i.e. does not itself reduces under
the same set of rules). Otherwise, in
\cite{ArrighiDiazcaroQPL09,ArrighiDiazcaroValironDCM11,BuirasDiazcaroJaskelioffLSFA11,ValironDCM10,ValironQPL10},
type systems are set up instead, to forbid diverging terms (modulo
associativity and commutativity of ``+'', or AC) such as $Y_M$.
Finally, in
$\oalg$ \cite{VauxMSCS09} a restriction to positive scalars is proposed to solve the
problem. 

Since any of these techniques will solve the problem, in this paper we do not make a choice {\em a priori}. Instead, we show
that the  simulations between the calculi are correct,
providing a methodology general enough to work in a large variety of
restrictions on the languages.
Therefore, we do not force a specific method to make the
calculi consistent, leaving the choice to the reader.

The simulation theorems that we develop in this paper are correct in
a general untyped setting (and in fact trivially true when we simulate
a language with algebraic reduction with a language with algebraic
equality as remarked above),
but also true if one restricts the scalars to a semiring (as done in
\cite{VauxMSCS09}), 
or if we restrict the terms to any typed setting,
provided that the languages $\llinred$ and $\lalgred$ satisfy
subject reduction 
and that the CPS translations given in Section \ref{sec:sim} preserve typability.
We propose a notion of {\em language
fragments} to parameterise the simulation results. The definition of a
fragment is general enough to capture many settings: various typed
systems, but also the restrictions to a given set of terms such as the set
of strongly normalising terms modulo AC
or taking scalars from a semiring.

We define formally a fragment in the following way:
\begin{definition}\label{def:subset} 
  A \emph{fragment} $S$ of $\llinred$ (respectively $\lalgred$) is a subset of terms closed under $\toblinred$-reduction (respectively $\tobalgred$-reduction) together with the rewrite system inherited from $\llinred$ (respectively $\lalgred$).
\end{definition}

The definition of a fragment in the presence of algebraic equalities
should be treated carefully. Indeed, note that the algebraic
equalities are defined as $M\to^=N$ if and only if $M\to N$ or $N\to
M$. As a consequence, for any subset $S$ of terms closed under
$\to^=$-reduction, if $M$ is in $S$ then for any $N$ (in $S$ or not),
$M+N-N\in S$ since $M\to^= M+N-N$. We therefore need to define the algebraic
equality with respect to the particular subset of terms under consideration.

\begin{definition}\label{def:subeq}
  A \emph{fragment} $S$ of $\llineq$ (respectively $\lalgeq$) is a fragment of
  $\llinred$ (respectively $\lalgred$) together with an algebraic equality
  defined as $M\fragment{S}{\tolineq} N$ (respectively $M\fragment{S}{\toalgeq} N$) if and only if
  $M,N\in S$ and $N\tolinred M $ or $M\tolinred N$ (respectively $N\toalgred
  M $ or $M\toalgred
  N$). The $\beta$-reduction is not modified.
\end{definition}

\begin{convention}
  When referring to a fragment of $\llineq$ (respectively $\lalgeq$), we use $\tolineq$
  (respectively $\toalgeq$) instead of $\fragment{S}{\tolineq}$ (respectively $\fragment{S}{\toalgeq}$) for the
  restricted rewrite system when the fragment under consideration is clear.
\end{convention}

\section{Simulations}\label{sec:sim}

The core of the paper is concerned with the mutual simulations of the
four languages. 
The first class of problems relates algebraic reduction with algebraic
equality. While simulating a language with algebraic reduction using a
language with algebraic equality is not difficult, going in
the opposite direction is not possible in general. Indeed, while $0
=_\ell Y_M - Y_M \tobv Y_M + M - Y_M =_\ell M$ in
$\llineq$ (where $Y_M=(\lambda x\,(M+(x)~x))~\lambda x\,(M+(x)~x)$),
it is impossible to reduce $0$ reduce to $M$ in $\llinred$, because
$0$ does not appear on the left-hand-side of any rule in
Figure~\ref{tab:RW}. 
In this section, we show that a fragment
of a language with algebraic equality can be simulated by the
corresponding fragment with algebraic reduction provided that the
latter is {\em confluent} (Theorems~\ref{thm:eqred} and~\ref{thm:eqred2}).

The second class of problems is concerned with call-by-value and
call-by-name. In this paper, the simulations of call-by-name by
call-by-value and its reverse are treated using continuation passing
style (CPS), extending the  techniques described
in~\cite{FischerSIGPLAN72,PlotkinTCS75} to the algebraic case
(Theorems~\ref{th:sim}, \ref{th:simeq}, \ref{th:sim2} and~\ref{th:sim2eq}).

The results are summarised in Figure~\ref{fig:relation}. Solid arrows
correspond to results where no particular hypothesis on the language
is made. Dotted arrows correspond to results where confluence is
required. Dashed arrows correspond to results where strong normalisation is required.

\subsection{Algebraic reduction versus algebraic equality}\label{sec:redeq}

As the relation $\toblinred$ is contained in $\toblineq$ and the
relation $\tobalgred$ is contained in $\tobalgeq$, the first
simulation theorems are trivial.

\begin{theorem}\label{th:redeq} For any term $M$ 
  if $M \tobalgred N$, then $M\tobalgeq N$.
  \qed
\end{theorem}

\begin{theorem}\label{th:redeq2} For any term $M$ 
  if $M \toblinred N$, then $M\toblineq N$. 
  \qed
\end{theorem}

The simulations going in the other direction are only valid in the
presence of confluence. In the following two theorems, the algebraic
equality is defined with respect to the considered fragment (see
Definition~\ref{def:subeq}.)  

\begin{theorem}\label{thm:eqred}
  For any term $M$ in a confluent fragment of $\llinred$, if $M\stoblineq V$, then there exists $V'$ such that $M\stoblinred V'$ and $V\stolineq V'$.
\end{theorem}
\begin{proof}
  By induction on the length of the reduction. The proof mainly relies
  on the confluence of the fragment. \App{eqred}.
\end{proof}

\begin{theorem}\label{thm:eqred2}
  For any term $M$ in a confluent fragment of $\lalgred$, if $M\stobalgeq V$, then there exists $V'$ such that $M\stobalgred V'$ and $V\stoalgeq V'$.
\end{theorem}
\begin{proof}
  Similar to the previous theorem.
\end{proof}

\subsection{Call-by-name simulates call-by-value}\label{subsec:lintoalg}

To prove the simulation of $\llinred$ in $\lalgred$ and the
simulation of $\llineq$ in $\lalgeq$, we introduce an
algebraic extension of the continuation passing style translation used to
prove that call-by-name simulates call-by-value in the regular
$\lambda$-calculus \cite{PlotkinTCS75}.

Let ~$\wt{\cdot}$ be the following encoding of terms from $\llinred/\llineq$ to $\lalgred/\lalgeq$, where $f,g$ and $h$ are fresh variables.
\[
  \begin{array}{rcl@{\hspace{1cm}}rcl}
    \wt{x} &=& \cont x,&
    \wt{0} &=& 0,\\
    \wt{\lambda x\,M} & =&\lambda f\,(f)~\lambda x\,\wt M,&
    \wt{(M)~N} &=& \tappfgh{\wt M}{\wt N},\\
    \wt{\alpha.M} &=&\lambda f\,(\alpha.\wt M)~f,&
    \wt{M+N}&=&\lambda f\,(\wt M +\wt N)~f.
  \end{array}
\]

Note that the CPS translation for the addition ($M+N$) is defined in a
non-standard manner. Indeed, if it were, say, a pairing construct, in
call-by-name, the term $M+N$ would always be a value, whereas in
call-by-value it would not necessarily be (for example if $M$ or $N$
reduces). One would therefore expect the CPS translation of $M+N$ to be
$\lambda f\,(\wt{M})\lambda x\,(\wt{N})\lambda y\,(f)(x+y)$ (similarly
as it is done for the application)
In the languages we are considering, however, if $M$ or $N$ reduces,
then $M+N$ reduces, even in a call-by-name setting. Thus the
definition we instead pick.

Let $\Psi$ be the encoding of base terms defined by
$\Psi(x) = x$,
$\Psi(\lambda x\,M) =\lambda x\,\wt M$,
This encoding is compatible with substitution:
\begin{lemma}\label{lem:substitution2}
  $\wt{M[x:=B]}=\wt{M}[x:=\Psi(B)]$ with $B$ a base term.
\end{lemma}
\begin{proof}
  By induction on $M$. \App{substitution2}.
\end{proof}

We also define a convenient infix operation ($:$) capturing the
behaviour of translated terms. For example, if $B$ is a base term,
{\em i.e.}\ a variable or an abstraction, then its translation into
$\lalgred$ is $\wt{B}=\cont{\Psi(B)}$. If we apply this translated term
to a certain $K$, we obtain $(\cont{\Psi(B)})~K\tobalgred
(K)~\Psi(B)$. We define $B:K=(K)~\Psi(B)$ and get that
$(\wt{B})~K\tobalgred B:K$. 

\begin{definition}
  Let $(:)$ be
  the infix binary operation defined by:
  \[
    \begin{array}{r@{\,=\,}l}
      0:K&0\\
      B:K&(K)~\Psi(B)\\
      \alpha.M:K&\alpha.(M:K)\\
      M+N:K&M:K+N:K\\
    \end{array}~~~
    \begin{array}{r@{\,=\,}l}
      (0)~N:K&0\\
      (B)~N:K&N:\lambda f\,((\Psi(B))~f)~K\\
      (\alpha.M)~N:K&\alpha.(M)~N:K \\
      (M+N)~L:K&((M)~L+(N)~L):K \\
      ((M)\,N)\,L:K&(M)\,N{:}\lambda\,g\,(\wt{L})\,\lambda\,h\,((g)\,h)\,K \\
    \end{array}
  \]
\end{definition}

Using this encoding, we can simulate $\llinred$ with $\lalgred$, as
formalised in the following theorem. The proof of this theorem is given later in this section.

\begin{theorem}[Simulation]\label{th:sim}
  For any term $M$ and variable $k$,
  if $M\stoblinred V$ where $V$ is a value,
  then $(\wt M)~k \stobalgred V:k$.
\end{theorem}

\begin{example}
  \label{ex:1}
  For any terms $M$ and $N$, let $\pair M N := \lambda y\,((y)\,M)\,N$.  Let $\textsf{copy}=\lambda x\, \pair x x$,  and let $B_1=\lambda x\, N_1$ and $B_2=\lambda x\, N_2$ be two base terms. Then $(\textsf{copy})~(B_1+B_2) \stoblinred
  \pair {B_1} {B_1} + \pair {B_2} {B_2}$ and $(\textsf{copy})~(B_1+B_2)\stobalgred
  \pair{B_1+B_2}{B_1+B_2}$.

  We consider the simulation $\llinred$ to $\lalgred$.
  \[
    \wt{(\textsf{copy})~(B_1+B_2) }=
    \lambda f\, (\wt{\textsf{copy}})\,
    \lambda g\, (\wt{B_1+B_2})\lambda h\, ((g)\,h)\,f\ ,
  \]
  where
  $\wt{\textsf{copy}}=
  \lambda f\, (f)\,\lambda x\, \wt{\pair x x}$,
  with
  $\wt{\pair M N}=
  \lambda f\, (f)\, \Psi(\pair M N)$,
  and $\wt{B_1+B_2}=
  \lambda f\, (\wt{B_1}+\wt{B_2})\,f$, with 
  $\wt{B_1}=
  \lambda g\, (g)\,\Psi(B_1)$.

  \begin{align*}
    (\wt{(\textsf{copy})~(B_1+B_2)})\,k
    &\tobalgred
    (\wt{\textsf{copy}})\,
    \lambda g\, (\wt{B_1+B_2})\lambda h\,((g)\,h)\,k
    \\
    &=
    (\lambda f\, (f)\,\lambda x\, \wt{\pair x x})\,
    \lambda g\, (\wt{B_1+B_2})\lambda h\, ((g)\,h)\,k
    \\
    &\tobalgred
    (\lambda g\, (\wt{B_1+B_2})\, \lambda h\, ((g)\,h)\,k)\,
    \lambda x\, \wt{\pair x x}
    \\
    &\tobalgred
    (\wt{B_1+B_2})\,\lambda h\, ((\lambda x\, \wt{\pair x x})\,h)\,k
    \\
    &\tobalgred
    (\wt B_1  + \wt B_2)\,
    \,\lambda h\, ((\lambda x\, \wt{\pair x x})\,h)\,k\\
    &\tobalgred
    (\wt B_1)\,
    \,\lambda h\, ((\lambda x\, \wt{\pair x x})\,h)\,k
    +
    (\wt B_2)\,
    \,\lambda h\, ((\lambda x\, \wt{\pair x x})\,h)\,k
    \\
    &\stobalgred
    (\lambda h\, ((\lambda x\, \wt{\pair x x})\,h)\,k)\, \Psi(B_1)
    +
    (\lambda h\, ((\lambda x\, \wt{\pair x x})\,h)\,k)\, \Psi(B_2)
    \\
    &\stobalgred
    ((\lambda x\, \wt{\pair x x})\,\Psi(B_1))\,k 
    +
    ((\lambda x\, \wt{\pair x x})\,\Psi(B_2))\,k 
    \\
    &\stobalgred
    (\wt{\pair x x}[x:= \Psi(B_1)])\,k 
    +
    (\wt{\pair x x}[x:= \Psi(B_2)])\,k  
    \\
    &\stobalgred
    (\wt{\pair {B_1}{B_1}})\,k 
    +
    (\wt{\pair {B_2}{B_2}})\,k
    \quad \textrm{(Lemma \ref{lem:substitution2})}
    \\
    &\stobalgred
    (k) \, \Psi({\pair {B_1}{B_1}})
    +
    (k) \, \Psi({\pair {B_2}{B_2}})\\
    &=
    ({\pair {B_1}{B_1}} + {\pair {B_2}{B_2}}):k
  \end{align*}
\end{example}

Similarly, one can relate fragments of $\lalgeq$ to fragments of
$\llineq$ as follows.

\begin{theorem}[Simulation]\label{th:simeq}
  For any two fragments $S_{\it \ell}$ of $\llineq$ and $S_{\it a}$ of $\lalgeq$ and variable $k$, such that $\forall M\in S_{\it \ell}$, $(\wt M)~k\in S_{\it a}$, and for any term $M$ in $S_{\it \ell}$, if $M\stoblineq V$ where $V$ is a value, then $(\wt M)~k \stobalgeq V:k$.
\end{theorem}
The proof of Theorem~\ref{th:simeq} is detailed later in this section.

\begin{remark}
  As we already noted several times in this paper, without restricting
  the languages, Theorem~\ref{th:simeq} would be trivial. If any term
  reduce to any other one, the desired reduction would of course be
  valid without restriction. This theorem shows that if the calculi
  are restricted to fragments, the result is still true.
  One example of such fragments is found by taking the restriction of
  scalars to non-negative elements, as in \cite{VauxMSCS09}.
\end{remark}

Once a term is encoded it can be reduced either by $\stobalgred$ or by
$\stoblinred$ (respectively $\stobalgeq$ or $\stoblineq$) without
distinction, and still obtain the same result. 
We state this fact as a corollary:

\begin{Corollary}[Indifference]\label{cor:ind}~
  \begin{enumerate}
    \item For any term $M$ and variable $k$, if $M\stoblinred V$ where $V$ is a value, then 
      \[ (\wt M)~k \stoblinred V:k \]
    \item For any  fragment $S$ of $\llineq$ and variable $k$, such that $\forall M\in S$, $(\wt M)~k\in S$, and for any term
      $M$ in $S$, if $M\stoblineq V$ where $V$ is a value, then 
      \[ (\wt M)~k \stoblineq V:k \]
  \end{enumerate}
\end{Corollary}
\begin{proof}
  It suffices to check that in the proofs of Theorems~\ref{th:sim}
  and~\ref{th:simeq} all the reductions $\stobalgred$
  are done by rules common to both languages.
\end{proof}

\begin{example}\label{ex:1-indifference}
  Note that in Example~\ref{ex:1} one could have as well rewritten with $\toblinred$, which illustrates the indifference property (Corollary \ref{cor:ind}).
\end{example}

We now proceed to prove Theorems~\ref{th:sim} and \ref{th:simeq}. 
We extend the proof in \cite{PlotkinTCS75} to the algebraic case.

\begin{lemma}\label{lem:lemma2}
  If $K$ is a base term, then for any term $M$, $(\wt{M})~K\stobalgred M:K$.
\end{lemma}
\begin{proof}
  By structural induction on $M$. \App{lemma2}.
\end{proof}

The following lemma and corollary state that the $(:)$ operation
preserves reduction.
\begin{lemma}\label{lem:lemma3alg}
  If $M\tolinred N$ then for all $K$ base term, $M:K\stoalgred N:K$.
\end{lemma}
\begin{proof}
  Case by case on the rules $\tolinred$. \App{lemma3alg}.
\end{proof}

\begin{lemma}\label{lem:lemma3}
  If $M\toblinred N$ then for all $K$ base term, $M:K\stobalgred N:K$.
\end{lemma}
\begin{proof}
  Case by case on the rules $\toblinred$. \App{lemma3}.
\end{proof}

\begin{corollary}\label{cor:lemma3}
  If $M\toblineq N$ then for all $K$ base terms, $M:K\stobalgeq N:K$.
\end{corollary}
\begin{proof}
  By case distinction. If $M\toblinred N$, then by
  Lemma~\ref{lem:lemma3}, $M:K\stobalgred N:K$, which implies
  $M:K\stobalgeq N:K$.  If $N\tolinred M$, then by
  Lemma~\ref{lem:lemma3alg}, $N:K\stoalgred M:K$, which also implies
  $M:K\stobalgeq N:K$.
\end{proof}

Now the proofs of Theorems~\ref{th:sim} and~\ref{th:simeq} go as follows.

\begin{proof}[\bf Proof of Theorem~\ref{th:sim}] 
  Using Lemmas~\ref{lem:lemma2} and \ref{lem:lemma3},
  we can deduce that the reduction
  $
  (\wt{M})~k\stobalgred M:k\stobalgred V:k
  $ happens.
\end{proof}

\begin{proof}[\bf Proof of Theorem~\ref{th:simeq}] From
  Lemma~\ref{lem:lemma2}, $(\wt{M})~k \stobalgred M:k$, hence we also have $(\wt{M})~k\stobalgeq M:k$.
  From Corollary~\ref{cor:lemma3}, this latter term $\stobalgeq$-reduces to $V:k$.
  Note that since $(\wt{M})~k\in S_{\it a}$,
  $M:k$ is also in $S_{\ell}$ due to the closeness under $\toalgeq$ of
  $S_{\it a}$. The same applies to $M:k$, thus also to
  $V:k$.
\end{proof}

\subsection{Completeness of the call-by-value to call-by-name simulation}\label{subsec:lintoalg-completeness}

We show that the converse of Theorem \ref{th:sim} is also true:
\begin{theorem}[Completeness]
  \label{thm:completeness} For any term $M$ and variable $k$, if $\wt Mk\stobalgred V:k$ then $M\stoblinred V$.
\end{theorem}
To prove it, we define an inverse translation and show that it preserves
reductions. First, we need to characterise the structure of
the encoded terms. We define a subset of $\lalgred/\lalgeq$ which contains
the image of the translation and is closed under $\tobalgred$ reductions
with the following grammar:

\[
  \begin{array}{rcll}
    C & ::= & (K)~B \mid ((B_{1})~B_{2})~K \mid (T)~K & \mbox{(base computations)}\\
    D & ::= & C\mid0\mid\alpha.D\mid D_{1}+D_{2} & \mbox{(computation combinations)}\\
    \\
    S & ::= & \lambda k\,C & \mbox{(base suspensions)}\\
    T & ::= & S\mid0\mid\alpha.T\mid T_{1}+T_{2} & \mbox{(suspension combinations)}\\
    \\
    K & ::= & k \mid \lambda b\,((B)~b)~K \mid \lambda{b_{1}}\,(T)~\lambda{b_{2}}\,((b_{1})~b_{2})~K & \mbox{(continuations)}\\
    \\
    B & ::= & x \mid \lambda x\,S & \mbox{(CPS-values)}
  \end{array}
\]

There are four main categories of terms: \emph{computations}, \emph{suspensions},
\emph{continuations}, and \emph{CPS-values}. We distinguish base computations
$C$ from linear combinations of computations $D$, as well as base
suspensions $S$ from linear combinations of suspensions $T$. The
translation $\wt M$ gives a term of the class $T$, while
$(\wt M)~k$ and $M:K$ are of class $D$. One can easily check
that each of the classes $D$, $T$, $K$ and $B$ is closed under $\tobalgred$
reductions.

For convenience, in this section we put some restrictions on the names of the variables in the
grammar. The variable name $k$ that appears in the class $K$ is the same as the one used in suspensions of the form $\lambda k\,C$.
It cannot appear as a variable name in any other term. This is to
agree with the requirement of freshness that we mentioned above. The
same applies for the variables $b$, $b_{1}$ and $b_{2}$: they cannot
appear (free) in any sub-term. In particular, these restrictions ensure
that the grammar \emph{for each category} is unambiguous. The three kinds
of variables ($x$, $k$ and $b$) play different roles, which is
why we distinguish them using different names.

Computations are the terms that simulate the steps of the reductions,
hence the name. They are the only terms that contain top-level applications,
so they are the only terms that can $\beta$-reduce. In fact, notice
that the arguments in applications are always base values. This shows
a simple alternative proof for Corollary \ref{cor:ind}.

We define the inverse translation using the following four functions,
corresponding to each of the four main categories in the grammar. These functions are well-defined because the grammar for each category
is unambiguous.

\[
  \begin{array}{rclrcl}
    \overline{(K)~B} & = & \underline{K}[\psi(B)] & \sigma(\lambda k\,C) & = & \overline{C}\\
    \overline{((B_{1})~B_{2})~K} & = & \underline{K}[(\psi(B_{1}))~\psi(B_{2})] & \sigma(0) & = & 0\\
    \overline{(T)~K} & = & \underline{K}[\sigma(T)] & \sigma(\alpha.T) & = & \alpha.\sigma(T)\\
    \overline{0} & = & 0 & \sigma(T_{1}+T_{2}) & = & \sigma(T_{1})+\sigma(T_{2})\\
    \overline{\alpha.D} & = & \alpha.\overline{D}\\
    \overline{D_{1}+D_{2}} & = & \overline{D_{1}}+\overline{D_{2}}\\
    &  &  & \underline{k}[M] & = & M\\
    \psi(x) & = & x & \underline{\lambda b\,((B)~b)~K}[M] & = & \underline{K}[(\psi(B))~M]\\
    \psi(\lambda x\,S) & = & \lambda x\,\sigma(S) & \underline{\lambda{b_{1}}\,(T)~\lambda{b_{2}}\,((b_{1})~b_{2})~K}[M] & = & \underline{K}[(M)~\sigma(T)]
  \end{array}
\]

To prove the completeness of the simulation we need
several technical lemmas.
The first two lemmas state that the translation defined
above is in fact an inverse.

\begin{lemma}
  \label{lem:inverse-term} For any term $M$, $\overline{(\wt M)~k}=M$.
\end{lemma}
\begin{proof}
  We have $\overline{(\wt M)~k}=\underline{k}[\sigma(\wt M)]=\sigma(\wt M)$
  so we have to show that $\sigma(\wt M)=M$ for all $M$. The
  proof follows by induction on the structure of $M$. \App{inverse-term}.
\end{proof}

In general, $\overline{M:k}\neq M$. Although it would be true for
a classical translation, it does not hold in the algebraic case. Specifically,
we have $(\alpha.M)L:k=\alpha.((M)~L):k$ and $(M+N)L:k=(M)~L+(N)~L:K$, so
the translation is not injective. However it is still true for values.

\begin{lemma}
  \label{lem:inverse-value} For any value $V$, $\overline{V:k}=V$.
\end{lemma}
\begin{proof}
  By induction on the structure of $V$. \App{inverse-value}.
\end{proof}

The last lemma (Lemma~\ref{lem:inverse-step}) states that the inverse translation preserves reductions.

\begin{lemma}
  \label{lem:inverse-step} For any computation $D$, if $D\tobalgred D'$
  then $\overline{D}\stoblinred\overline{D'}$. Also, if $D\toalgred D'$
  then $\overline{D}\stolinred\overline{D'}$.
\end{lemma}
\begin{proof}
  The proof of this lemma follows by induction on the reduction rules.
  It uses the following several intermediary results.
  \begin{itemize}
    \item 
      The following equalities hold:
      \begin{enumerate}
	\item $\psi(B_{1})[x:=\psi(B)]=\psi(B_{1}[x:=B])$
	\item $\sigma(T)[x:=\psi(B)]=\sigma(T[x:=B])$
	\item $\overline{C}[x:=\psi(B)]=\overline{C[x:=B]}$
	\item
	  $\underline{K}[M][x:=\psi(B)]=\underline{K[x:=B]}[M[x:=\psi(B)]]$
      \end{enumerate}
    \item 
      For all terms $M$ and continuations
      $K_{1}$ and $K_{2}$, 
      $
      \underline{K_{1}}[\underline{K_{2}}[M]]\!=\!\underline{K_{2}[k:=K_{1}]}[M]
      $.
    \item For all $K$ and $C$,
      $\underline{K}[\overline{C}]=\overline{C[k:=K]}$ (using the
      previous item)
    \item For any continuation $K$ and term
      $M$, if $M\toblinred M'$, then
      $\underline{K}[M]\toblinred\underline{K}[M']$.
    \item 
      The following relations hold (by induction on $K$, using the
      previous item)
      \begin{itemize}
	\item $\underline{K}[M_{1}+M_{2}]\stolinred\underline{K}[M_{1}]+\underline{K}[M_{2}]$
	\item $\underline{K}[\alpha.M]\stolinred\alpha.\underline{K}[M]$
	\item $\underline{K}[0]\stolinred0$
      \end{itemize}
    \item For any suspension $T$, if $T\toalgred T'$
      then $\sigma(T)\tolinred\sigma(T')$.  
  \end{itemize}
  \App{inverse-step}.
\end{proof}

With these we can prove the completeness theorem.

\begin{proof}[\bf Proof of Theorem \ref{thm:completeness}]
  Using Lemma \ref{lem:inverse-step} for each step of the reduction,
  we get that $\overline{(\wt M)~k}\stoblinred\overline{V:k}$. By Lemma
  \ref{lem:inverse-term} and Lemma \ref{lem:inverse-value}, this implies
  $M\stoblinred V$.
\end{proof}

\subsection{Call-by-value simulates call-by-name}\label{subsec:algtolin}

To state that $\llinred$ simulates $\lalgred$, we again use an algebraic
extension of the continuation passing style encoding of ~\cite{PlotkinTCS75}. Let ~$\cps{\cdot}$ be the following encoding of terms from $\lalgred$ to $\llinred$ where
$f,g$ and $h$ are fresh variables.
\[
  \begin{array}{rcl@{\hspace{1cm}}rcl}
    \cps{x} &=&x,&
    \cps{0} &=& \lambda f\, (0)~f,\\
    \cps{\lambda x\,M} & =&\lambda f\,(f)~\lambda x\,\cps M,&
    \cps{(M)~N} &=& \tappcps{\cps M}{\cps N},\\
    \cps{\alpha.M} &=& \lambda f\, (\alpha.\cps M)~f,&
    \cps{M+N}&=&\lambda f\,(\cps M +\cps N) ~f.
  \end{array}
\]
This encoding satisfies two useful properties.
\begin{lemma}\label{3:lem:base}
  For all terms $M$, the term $\cps{M}$ is a base term.\qed
\end{lemma}

\begin{lemma}\label{lem:substitution-cps}
  $\cps{M[x:=N]}=\cps{M}[x:=\cps N]$.
\end{lemma}
\begin{proof}
  By structural induction on $M$. \App{substitution-cps}.
\end{proof}

Let $\Phi$ be the encoding of abstractions defined by
$\Phi(\lambda x\,M)\,{=}$ $\lambda x\,\cps M$.
We keep the same notation ``:'' for the administrative infix operation capturing the behaviour of translated terms.

\begin{definition}
  Let $(:)$ be
  the infix binary operation defined by:
  \[
    \begin{array}{r@{\,=\,}l}
      0:K&0\\
      x:K&(x)~K\\
      \lambda\,x\,M:K&(K)~\Phi(\lambda\,x\,M)\\
      \alpha.M:K&\alpha.(M:K)\\
      M+N:K&M:K+N:K\\
    \end{array}~~~~
    \begin{array}{r@{\,=\,}l}
      (0)~N:K&0\\
      (x)~N:K&x:\lambda\,f\,((f)\,\cps{N})\,K\\
      (\lambda\,x\,M)~N:K&((\Phi(\lambda\,x\,M))~\cps N)~K \\
      (\alpha.M)~N:K&\alpha.(M)~N:K \\
      (M+N)~L:K&((M)~L+(N)~L):K \\
      ((M)\,N)\,L:K&(M)\,N{:}\lambda\,f\,((f)\,\cps{L})\,K
  \end{array}\]
\end{definition}
Simulation theorems, similar to Theorems~\ref{th:sim} and~\ref{th:simeq}, can be stated as follows.

\begin{theorem}[Simulation]\label{th:sim2}
  For any term $M$ and variable $k$, 
  if $M\stobalgred V$ where $V$ is a value, then $(\cps M)~k \stoblinred V:k$.
\end{theorem}

\begin{theorem}[Simulation]\label{th:sim2eq}\label{th:simeq2}
  For any two fragments $S_{\it a}$ of $\lalgeq$ and $S_{\it \ell}$ of
  $\llineq$ and variable $k$, such that $\forall M\in S_{\it a}$, $(\cps M)~k\in S_{\it \ell}$,
  and for any term $M$ in $S_{\it a}$, if $M\stobalgeq V$ where $V$ is
  a value, then $(\cps M)~k\stoblineq V:k$.
\end{theorem}

A result similar to Corollary~\ref{cor:ind} can also be
formulated. It is proven in a similar manner.

\begin{corollary}[Indifference]\label{cor:ind2}~
  \begin{enumerate}
    \item For any term $M$ and variable $k$, if $M\stobalgred V$ where $V$ is a value, then 
      \[(\cps M)~k\stobalgred V:k\]
    \item For any  fragment $S$ of $\lalgeq$ and variable $k$, such that $\forall M\in S$, $(\cps M)~k\in S$, 
      and for any term $M$ in $S$, if $M\stobalgeq V$ where $V$ is a value, then 
      \[(\cps M)~k\stobalgeq V:k\]
  \end{enumerate}
\end{corollary}

Before moving to the description of the proofs of
Theorems~\ref{th:sim2} and~\ref{th:sim2eq}, let us consider an
example.

\begin{example}\label{ex:2}
  We illustrate Theorem~\ref{th:sim2} using the term $(\textsf{copy})~(B_1+B_2)$ of Example~\ref{ex:1} which reduces to $\pair {B_1} {B_1} +\pair {B_2} {B_2}$ in $\llinred$ and to $\pair{B_1+B_2}{B_1+B_2}$ in $\lalgred$.
  The translation $\cps{(\textsf{copy})~(B_1+B_2) }$ is the term
  $\lambda f\, (\cps{\textsf{copy}})\,
  \lambda g\, ((g) \, \cps{B_1+B_2})\, f$,
  where
  $\cps{\textsf{copy}}$ is $\lambda f\, (f)\,\lambda x\, \cps{\pair x x}$.
  $\cps{\pair M N}$ is
  $\lambda f\, (f)\, \Phi(\pair M N)$, 
  $\cps{B_1+B_2}$ is 
  $\lambda f\, (\cps{B_1}+\cps{B_2})\,f$
  and
  $\cps{B_1}$ is $\lambda g\, (g)\,\Phi(B_1)$

  \[
    \begin{array}{rll}
      (\cps{(\textsf{copy})~(B_1+B_2)})\,k&\toblinred&
      (\cps{\textsf{copy}})\,
      \lambda g\, ((g) \, \cps{B_1+B_2})\,k
      \\
      &=&
      (\lambda f\, (f)\,\lambda x\, \cps{\pair x x})\,
      \lambda g\, ((g) \, \cps{B_1+B_2})\,k
      \\
      &\toblinred&
      (\lambda g\, ((g) \, \cps{B_1+B_2})\,k)\,
      \lambda x\, \cps{\pair x x}
      \\
      &\toblinred&
      ((\lambda x\, \cps{\pair x x}
      ) \, \cps{B_1+B_2})\,k
      \\
      \textrm{(Lemma \ref{3:lem:base})}
      &\toblinred&
      (\cps{\pair x x}[x:=  \cps{B_1+B_2}]
      )\,k
      \\
      \textrm{(Lemma \ref{lem:substitution-cps})}
      &=&
      (\cps{\pair {B_1+B_2} {B_1+B_2}})\,k
      \\
      &\toblinred&
      (k) \, \Phi({\pair {B_1+B_2} {B_1+B_2}}) 
      \\
      &=&
      {\pair {B_1+B_2} {B_1+B_2}}:k
    \end{array}
  \]
\end{example}

In Section~\ref{subsec:lintoalg}, the proofs of the simulations theorems
were performed using two intermediate results, as follows (the term $K$ is taken as a base term).
\begin{enumerate}
  \item Prove that $(\wt{M})\,K \stobalgred M:K$;
  \item prove that if $M\toblinred N$ then $M:K\stobalgred N:K$.
\end{enumerate}
For the simulation theorems of the present section, we use a similar
procedure. 
The three lemmas needed for the proof of the simulation theorems now
read as follow.

\begin{lemma}\label{lem:lemma2cps}
  If $K$ is a base term,  then for every term $M$ we have
  $(\cps{M})~K\stoblinred M:K$.
\end{lemma}
\begin{proof}
  By structural induction on $M$. \App{lemma2cps}.
\end{proof}

\begin{lemma}\label{lem:lemma3cps}
  If $M\tobalgred N$ then for every $K$ base term we have $M:K\stoblinred N:K$.
\end{lemma}
\begin{proof}
  Case by case on the rules of $\lambda_{\textit{alg}}$. \App{lemma3cps}.
\end{proof}

We are now ready to prove the simulation theorems. As advertised,
these proofs reflect the exact same structures of the proofs of
Theorems~\ref{th:sim} and~\ref{th:simeq}.

\begin{proof}[\bf Proof of Theorem~\ref{th:sim2}]
  From Lemmas~\ref{lem:lemma2cps} and \ref{lem:lemma3cps}, we have the
  following reduction.
  $
  (\cps{M})~k\stoblinred M:k\stoblinred V:k
  $.
\end{proof}

\begin{proof}[\bf Proof of Theorem~\ref{th:simeq2}] 
  From Lemma~\ref{lem:lemma2cps}, $(\cps{M})~k \stoblinred M:k$, hence we also have $(\cps{M})~k\stoblineq M:k$. 
  A result equivalent to Corollary~\ref{cor:lemma3} can be shown as easily:
  if $M\toalgeq N$ then for all base terms $K$, $M:K\stolineq N:K$.
  This entails that $M:k\stoblineq V:k$.
  Note that since $(\cps{M})~k\in S_{\ell}$, $M:k$ is also in $S_{\ell}$ due to the closeness under $\tolineq$ of
  $S_{\it \ell}$. The same argument applies to $M:k$, thus also to
  $V:k$.
\end{proof}

\subsection{Completeness of the call-by-name to call-by-value
simulation}\label{subsec:algtolin-completeness}

We use the same procedure as in Section \ref{subsec:lintoalg-completeness} to show
that the translation is also complete.

\begin{theorem}[Completeness]
  \label{thm:completeness-a} For any term $M$ and variable $k$, if $(\cps M)~k\stoblinred V:k$ then
  $M\stobalgred V$.
\end{theorem}

The adjustments we have to make are the same as in the classical case: we treat variables and applications differently. Bellow is the grammar of the target language. It is closed under $\toblinred$
reductions.

\[
  \begin{array}{rcll}
    C & ::= & (K)~B \mid ((B)~S)~K \mid (T)~K & \mbox{(base computations)}\\
    D & ::= & C \mid 0 \mid \alpha.D \mid D_{1}+D_{2} & \mbox{(computation combinations)}\\\\
    S & ::= & x \mid \lambda k\,C & \mbox{(base suspensions)}\\
    T & ::= & S \mid 0 \mid \alpha.T \mid T_{1}+T_{2} & \mbox{(suspension combinations)}\\\\
    K & ::= & k \mid \lambda b\,((b)~S)~K & \mbox{(continuations)}\\\\
    B & ::= & \lambda x\,S & \mbox{(CPS-values)}
  \end{array}
\]

Notice how $x$ is now considered a suspension, not a CPS-value. This
is because $x$ is replaced by a suspension after beta-reducing a
term of the form $((\lambda x\,S)~S)~K$. This is the main difference between
the call-by-name and call-by-value CPS simulations. Apart from that,
it satisfies the same properties.

We define the inverse translation using the following four functions.

\[
  \begin{array}{rclrcl}
    \overline{(K)~B} & = & \underline{K}[\phi(B)] & \sigma(x) & = & x\\
    \overline{((B)~S)~K} & = & \underline{K}[(\phi(B))~\sigma(S)] & \sigma(\lambda k\,C) & = & \overline{C}\\
    \overline{(T)~K} & = & \underline{K}[\sigma(T)] & \sigma(0) & = & 0\\
    \overline{0} & = & 0 & \sigma(\alpha.T) & = & \alpha.\sigma(T)\\
    \overline{\alpha.D} & = & \alpha.\overline{D} & \sigma(T_{1}+T_{2}) & = & \sigma(T_{1})+\sigma(T_{2})\\
    \overline{D_{1}+D_{2}} & = & \overline{D_{1}}+\overline{D_{2}}\\
    &  &  & \underline{k}[M] & = & M\\
    \phi(\lambda x\,S) & = & \lambda x\,\sigma(S) & \underline{\lambda b\,((b)~S)~K}[M] & = & \underline{K}[(M)~\sigma(S)]
  \end{array}
\]

To prove the completeness of the simulation we need analogous lemmas.
Their proofs are similar, but we need to account for the changes mentioned
above. 

\begin{lemma}
  \label{lem:inverse-term-a} For any term $M$, $\overline{(\cps M)~k}=M$.
\end{lemma}
\begin{proof}
  By induction on $M$. \App{inverse-term-a}.
\end{proof}

\begin{lemma}
  \label{lem:inverse-value-a} For any value $V$, $\overline{V:k}=V$.
\end{lemma}
\begin{proof}
  By induction on $V$. \App{inverse-value-a}.
\end{proof}

\begin{lemma}
  \label{lem:inverse-step-a} For any computation $D$, if $D\toblinred D'$
  then $\overline{D}\stobalgred\overline{D'}$. Also, if $D\tolinred D'$
  then $\overline{D}\stoalgred\overline{D'}$.
\end{lemma}
\begin{proof}
  The proof of this lemma is very similar to the one of
  Lemma~\ref{lem:inverse-step}. It follows by induction, involving several
  intermediary results similar to the ones in the proof of
  Lemma~\ref{lem:inverse-step}, where: $\phi$ is replaced with
  $\sigma$ and $\sigma$ with $\psi$ in the four first equalities; then:
  $\stolinred$ is replaced with $\stoalgred$ and $\tolinred$ is
  replaced with $\toalgred$. In other words:
  \begin{itemize}
    \item  The following equalities hold.
      \begin{enumerate}
	\item {$\phi(B)[x:=\sigma(S)]=\phi(B[x:=S])$}
	\item $\sigma(T)[x:=\sigma(S)]=\sigma(T[x:=S])$
	\item {$\overline{C}[x:=\sigma(S)]=\overline{C[x:=S]}$}
	\item $\underline{K}[M][x:=\sigma(S)]=\underline{K[x:=S]}[M[x:=\sigma(S)]]$
      \end{enumerate}
    \item For all terms $M$ and continuations
      $K_{1}$ and $K_{2}$, 
      $\underline{K_{1}}[\underline{K_{2}}[M]]=\underline{K_{2}[k:=K_{1}]}[M]$.
    \item For all $K$ and
      $C$, $\underline{K}[\overline{C}]=\overline{C[k:=K]}$.
    \item For any continuation $K$ and term
      $M$, if $M\tobalgred M'$ then $\underline{K}[M]\tobalgred\underline{K}[M']$.
    \item For any continuation $K$, scalar $\alpha$ and terms $M$, $M_1$
      and $M_2$, the following relations hold.
      \begin{itemize}
	\item $\underline{K}[M_{1}+M_{2}]\stoalgred\underline{K}[M_{1}]+\underline{K}[M_{2}]$
	\item $\underline{K}[\alpha.M]\stoalgred\alpha.\underline{K}[M]$
	\item $\underline{K}[0]\stoalgred0$
      \end{itemize}
    \item For any suspension $T$, if $T\tolinred T'$
      then $\sigma(T)\toalgred\sigma(T')$.
  \end{itemize}
  \App{inverse-step-a}.
\end{proof}

Finally, we can prove the completeness theorem using the previous lemmas.

\begin{proof}[\bf Proof of Theorem \ref{thm:completeness-a}]
  Using Lemma \ref{lem:inverse-step-a} for each step of the reduction,
  we get that $\overline{(\cps M)~k}\stobalgred\overline{V:k}$. By Lemma
  \ref{lem:inverse-term-a} and Lemma \ref{lem:inverse-value-a}, this
  implies $M\stobalgred V$.
\end{proof}

\subsection{The remaining simulations}\label{subsec:compose}

In Figure~\ref{fig:relation}, some arrows are missing, for example from \llinred to \lalgeq. We now show that the already existing arrows ``compose'' well. The first two simulations are $\lalgred\to\llineq$ and $\llinred\to\lalgeq$ and do not require confluence.
\begin{theorem}
  For any term $M$ and variable $k$,
  if $M\stoblinred V$ (respectively $M\stobalgred V$) where $V$ is a value, then $(\wt M)~k \stobalgeq V:k$ (respectively $(\cps M)~k \stoblineq V:k$).
\end{theorem}
\begin{proof}
  Given that $M\stoblinred V$, by Theorem~\ref{th:sim}, $(\wt M)~k\stobalgred V:k$, which by Theorem~\ref{th:redeq} implies $(\wt M)~k\stobalgeq V:k$.
  Analogously, given that $M\stobalgred V$, by Theorem~\ref{th:sim2}, we have
  $(\cps M)~k\stoblinred V:k$, which by
  Theorem~\ref{th:redeq2} implies that $(\cps M)~k$ $\stoblineq$ $V:k$.
\end{proof}

The other two simulations are $\lalgeq\to\llinred$ and
$\llineq\to\lalgred$ and they do require confluence.
\begin{theorem}
  For any term $M$ in a confluent fragment of $\llineq$
  (respectively $\lalgeq$) and variable $k$, if $M\stoblineq V$ (respectively $M\stobalgeq V$) then we have
  $(\wt M)~k \stobalgred V':k$ with  $V\stolineq V'$
  (respectively $(\cps M)~k \stoblinred V':k$ with  $V\stoalgeq
  V'$).
\end{theorem}
\begin{proof}
  Given that $M\stoblineq V$ and that $M$ is in a confluent fragment, Theorem~\ref{thm:eqred} states that $M\stoblinred V'$ with $V\stolineq V'$. In addition, Theorem~\ref{th:sim} states that $(\wt
  M)~k\stobalgred V':k$.  
  The other result is similar using
  Theorems~\ref{thm:eqred2} and \ref{th:sim2}. 
\end{proof}

Finally, we show that we can also compose with the
inverse translations to give the completeness of the remaining simulations.
This time however, the requirements of confluence are reversed.

\begin{theorem}
  For any term $M$ and variable $k$, if $(\wt M)~k\stobalgred V:k$ (respectively $(\cps M)~k\stoblinred V:k$)
  then for any fragment $S$ of $\llineq$ (respectively $\lalgeq$)
  such that $M\in S$, we have $M\stoblineq V'$ (respectively $M\stobalgeq V$) in $S$.\end{theorem}
\begin{proof}
  If $(\wt M)~k\stobalgred V:k$ then by Theorem \ref{thm:completeness}
  $M\stoblinred V$ which implies $M\stoblineq V$. Similarly if $(\cps M)k\stoblinred V:k$
  then by Theorem \ref{thm:completeness-a} $M\stobalgred V$ which
  implies $M\stobalgeq V$.\end{proof}

\begin{theorem}
  Let $S$ be a confluent fragment of $\lalgeq$ (respectively $\llineq$).
  For any term $M$ and variable $k$ such that $(\wt M)~k$ (respectively $(\cps M)~k$)
  is in $S$, if $\wt Mk\stobalgeq V:k$ (respectively $(\cps M)~k\stoblineq V:k$)
  then $M\stoblinred V'$ with $V\stolineq V'$ (respectively $M\stobalgred V'$
  with $V\stoalgeq V'$).\end{theorem}
\begin{proof}
  If $(\wt M)~k\stobalgeq V:k$ then by Theorem \ref{thm:eqred}
  $(\wt M)~k\stobalgred W$ with $V:k\stoalgeq W$. By Lemma \ref{lem:inverse-step},
  we get $\overline{(\wt M)~k}\stoblinred\overline{W}$
  with $\overline{V:k}\stolineq\overline{W}$, which by Lemmas \ref{lem:inverse-term}
  and \ref{lem:inverse-value} imply $M\stoblinred V':=\overline{W}$ with
  $V\stolineq V'$.\\
  Similarly, if $(\cps M)~k\stoblineq V:k$ then by Theorem \ref{thm:eqred2}
  $(\cps M)~k\stoblinred W$ with $V:k\stolineq W$. By Lemma \ref{lem:inverse-step-a},
  we get $\overline{(\cps M)~k}\stobalgred\overline{W}$
  with $\overline{V:k}\stoalgeq\overline{W}$, which by Lemmas \ref{lem:inverse-term-a}
  and \ref{lem:inverse-value-a} imply $M\stobalgred V':=\overline{W}$
  with $V\stoalgeq V'$.
\end{proof}

\section{Discussion and perspectives}\label{sec:concl}

\subsection{Simulating cloning}
As we discussed in the introduction, $\olin$ is a language whose
original purpose was to emulate quantum superpositions of states with
linear combinations of terms. However, we saw in Example \ref{ex:2}
that we can emulate the ``cloning'' operation
$(\textsf{copy})~(B_1+B_2) \stobalgred \pair {B_1+B_2} {B_1+B_2}$ in
$\olin$ using a CPS encoding. How does this relate to the no-cloning
theorem \cite{WoottersZurekNATURE82} stating that a quantum state
cannot be duplicated?

Since $\olin$ is a higher-order
language, a term both represents a quantum operator (i.e. a linear
map) and a state of the system (i.e. a vector in the space of states).
The choice of a call-by-value reduction strategy enforces this
philosophy: an application $(M)\,(N + L)$ is really $(M)\,N + (M)\,L$,
and there is no reduction under lambda's, making lambda-abstractions
correspond to ``pieces of code'' to be executed only when applied. So
the term $\lambda y\,(N+L)$ really stands for a piece of code that
would input a base vector $y$ and produce (possibly after some
process) a superposition $N+L$. But in itself, the lambda-abstraction
is a base vector -- it is not a linear combination. If we were to use
it, say as argument of $M=\lambda f\,\lambda x\,(f)\,(f)\,x$, it would
actuallly get duplicated.

On the contrary, the term $\lambda y\,N+\lambda y\,L$ is the linear
combination of two operators\,: one that inputs $y$ and produces $N$,
the other one that inputs $y$ and produces $L$. Fed to the same term
$M=\lambda f\,\lambda x\,(f)\,(f)\,x$, the distributivity of addition
over application would take precedence: the term $M$ behaves as a
linear operator and $(M)\,(\lambda y\,N+\lambda y\,L)$ is really
$(M)\,\lambda y\,N+(M)\,\lambda y\,L$.

We can see the same pattern in the term $(\textsf{copy})~(B_1+B_2)$\,:
the argument to the (linear) operator $\textsf{copy}$ is a linear
combination of $B_1$ and $B_2$, therefore in $\olin$ the term
$(\textsf{copy})~(B_1+B_2)$ really corresponds to $\pair{B_1}{B_1} +
\pair{B_2}{B_2}$ and not to $\pair{B_1+B_2}{B_1+B_2}$.  Along the CPS
transformation from $\lalgred$ to $\llinred$, recall from
Section~\ref{subsec:algtolin} that the argument $B_1+B_2$ is
transformed into $\lambda f\,(\cps{B_1} + \cps{B_2})\,f$. We are
therefore not anymore in presence of a linear combination, but of
a program that eventually produces a superposition. But this program
is not a superposition: it is a base state that will be fed unchanged
to an operator. In particular, if this operator is duplicating its
argument, the code $\lambda f\,(\cps{B_1} + \cps{B_2})\,f$ will be
duplicated. But instead of duplicating a superposition of terms, we
are really duplicating the description of a program eventually
producing a superposition.

\subsection{Conclusion}
In this paper we have shown the relation between two algebraic
$\lambda$-calculi, $\oalg$ and $\olin$, via four canonical languages.
These canonical algebraic $\lambda$-calculi account for all the
different choices we can make between call-by-value versus
call-by-name and algebraic reduction versus algebraic equality. We
showed how each language can simulate the other, by taking care of
marking where confluence was used.

This study opens the door to other questions.  The calculus $\oalg$
admits finiteness spaces as a model
\cite{EhrhardMSCS05,EhrhardLICS10}. What is the structure of the model
of the linear algebraic $\lambda$-calculus induced by the
continuation-passing style translation in finiteness spaces? The
algebraic lambda-calculus can be equipped with a differential
operator.  What is the corresponding operator in call-by-value through
the translation?  The linear-algebraic lambda-calculus can encode
quantum programs \cite{ArrighiDiazcaroValiron13}. Can this translation
help elucidate the relation between quantum computing and finiteness
spaces?

\paragraph*{Acknowledgements} 
We would like to thank Pablo Arrighi and Lionel Vaux for fruitful discussions and suggestions.

\bibliographystyle{alpha}
\bibliography{biblio}
\appendix

\setlist{topsep=0pt,parsep=0pt,partopsep=0pt,
  labelindent=0ex,labelwidth=1.5ex,labelsep=1.5ex,
leftmargin=3ex,itemindent=0ex}

\newenvironment{myenumerate}{\begin{enumerate}[labelindent=0ex,labelwidth=3ex,labelsep=1.5ex,leftmargin=4.5ex,itemindent=0ex]}{\end{enumerate}}

\begin{absolutelynopagebreak}

  \section{Detailed proofs}\label{sec:detailled-proofs}

  \subsection{Proof of Theorem~\ref{thm:LLINtoLIN}}\label{proof:LLINtoLIN}~

  \recap{Theorem}{thm:LLINtoLIN} 
  If $M$ is closed and strongly normalising in $\olin^*$ and $M\toblinred N$,
  then there exists $L$ such that $M\toolin^* L$ and $N\toolin^* L$.
\end{absolutelynopagebreak}
\begin{proof}
  We proceed by induction on the $\toblinred$ rewrite relation.
  The differences between $\olin^*$ rules and $\llinred$ rules are only in the conditions of rules $A_l$ and $A_r$, the three first factorisation rules and the context rule $\xi_{\lalin}$. Hence, if $M\toolin N$, we can just take $L=N$. So, it suffices to consider only these different rules, when they do not coincide with those in $\olin^*$. 
  \begin{myenumerate}
    \item\label{case:leftLin} $(M+N)~V\toblinred (M)~V+(N)~V$, with $M+N$ not normal in $\olin^*$ (it is closed by assumption). Let $L$ be the normal form in $\olin^*$ of $M+N$. Cases:
      \begin{itemize}
	\item $L=M'+N'$ with $M\stoolin M'$ and $N\stoolin N'$. Then we have
	  $
	  (M+N)~V\stoolin (M'+N')~V\toolin(M')~V+(N')~V
	  $,
	  and also
	  $
	  (M)~V+(N)~V\stoolin (M')~V+(N')~V
	  $.
	\item $L=M'$, with $M\stoolin M'$ and $N\stoolin 0$. Then we have
	  $
	  (M+N)~V\stoolin (M')~V
	  $,
	  and also
	  $
	  (M)~V+(N)~V\stoolin(M')~V+(0)~V\toolin(M')~V+0\toolin(M')~V
	  $.
	\item $L=(\alpha+\beta).L'$ with $M\stoolin\alpha.L'$ and $N\stoolin\beta.L'$. Then we have
	  $(M+N)~V$
	  $\stoolin((\alpha+\beta).L')~V\toolin(\alpha+\beta).(L')~V
	  $,
	  and also
	  $  
	  (M)~V+(N)~V \stoolin(\alpha.L')~V+(\beta.L')~V$ $\stoolin\alpha.(L')~V+\beta.(L')~V
	  \toolin(\alpha+\beta).(L')~V
	  $.
	\item Cases $L=(\alpha+1).L'$ with $M\stoolin\alpha.L'$ and $N\stoolin L'$, and $L=(1+1).L'$ with $M\stoolin L'$ and $N\stoolin L'$ are analogous to the previous case.
      \end{itemize}
    \item $(B)~(M+N)\toblinred(B)~M+(B)~N$, with $M+N$ not normal in $\olin^*$. This case is analogous to case~\ref{case:leftLin}.
    \item\label{case:leftScalarLin} $(\alpha.M)~V\toblinred \alpha.(M)~V$, with $M$ not normal in $\olin^*$. Let $M'$ be the normal form in $\olin^*$ of $M$. Then 
      $
      (\alpha.M)~V\stoolin(\alpha.M')~V\toolin\alpha.(M')~V
      $
      and
      $\alpha.(M)~V$
      $\stoolin\alpha.(M')~V$.

    \item $(B)~(\alpha.M)\toblinred\alpha.(B)~M$, with $M$ not normal in $\olin^*$. This case is analogous to case~\ref{case:leftScalarLin}.

    \item $(0)~V\toblinred 0$. Notice that $(0)~V\toolin 0$.

    \item $(B)~0\toblinred 0$. Notice that $(B)~0\toolin 0$.

    \item\label{thm:LLINtoLIN:it:fact} $\alpha.M+\beta.M\toblinred(\alpha+\beta).M$, when $M$ is not normal. Let $L$ be the normal form in $\olin^*$ of $M$. Then $\alpha.M+\beta.M\toolin^*\alpha.L+\beta.L\toolin(\alpha+\beta).L$, and $(\alpha+\beta).M\toolin^*(\alpha+\beta).L$.
    \item $\alpha.M+M\toblinred(\alpha+1).M$ and $M+M\toblinred(1+1).M$, when $M$ is not normal. Analogous to case \ref{thm:LLINtoLIN:it:fact}.
    \item $(V)~M\toblinred(V)~M'$, with $M\toblinred M'$. By the induction hypothesis, there exists $L$ such that $M\stoolin L$ and $M' \stoolin L$. Hence we have $(V)~M\stoolin(V)~L$ and also $(V)~M'\stoolin(V)~L$. \qedhere
  \end{myenumerate}
\end{proof}

\subsection{Proof of Theorem~\ref{thm:LINtoLLIN}}\label{proof:LINtoLLIN}~
\mynobreakpar
\recap{Theorem}{thm:LINtoLLIN}
If $M$ is closed and strongly normalising in $\olin^*$ and  $M\toolin N$, then there exists $L$ such that $M\stoblinred L$ and $N\stoblinred L$.\mynobreakpar
\begin{proof}
  We only need to verify the seven differing rules between the two languages.
  Notice that the normal form of a closed term is a value $V$.
  \begin{myenumerate}
    \item $(V_1+V_2)~L\toolin (V_1)~L+(V_2)~L$, with $V_1+V_2$ closed normal. 
      Let $L\toblinred W$, then we have
      $(V_1+V_2)~L\stoblinred (V_1+V_2)~W\toblinred (V_1)~W+(V_2)~W$,
      and also
      $(V_1)~W+(V_2)~W\stoblinred(V_1)~W+(V_2)~W$.
    \item $(L)~(V_1+V_2)\toolin(L)~V_1+(L)~V_2$. A value $V$ can be $0$ or a linear combination of base terms.
      \begin{itemize}
	\item Let $0$ be the normal form of $L$. Hence,
	  $(L)~(V_1+V_2)\stoblinred(0)~(V_1+V_2)\toblinred 0$
	  and
	  $(L)~V_1+(L)~V_2\stoblinred (0)~V_1+(0)~V_2\toblinred 0+0\toblinred 0$
	\item Let $\sum\limits_{i=1}^n\alpha_i.B_i$ be the normal form of $L$. Then
	  $(L)~(V_1+V_2)$
	  $\stoblinred$ $(\sum\limits_{i=1}^n\alpha_i.B_i)~(V_1+V_2)$
	  $\stoblinred$ $\sum\limits_{i=1}^n\alpha_i.(B_i)~(V_1+V_2)$
	  $\stoblinred$ $\sum\limits_{i=1}^n\alpha_i.((B_i)~V_1+(B_i)~V_2)$
	  $\stoblinred$ $\sum\limits_{i=1}^n\alpha_i.(B_i)$
	  $V_1+\sum\limits_{i=1}^n\alpha_i.(B_i)~V_2$,
	  and
	  $(L)~V_1+(L)~V_2$
	  $\stoblinred$ $(\sum\limits_{i=1}^n\alpha_i.B_i)~V_1+(\sum\limits_{i=1}^n\alpha_i.B_i)~V_2$
	  $\stoblinred$ $\sum\limits_{i=1}^n\alpha_i.(B_i)~V_1+\sum\limits_{i=1}^n\alpha_i.(B_i)~V_2$.
      \end{itemize}

    \item $(\alpha.V)~M\toolin \alpha.(V)~M$. Let $M\stoblinred W$, then 
      $(\alpha.V)~W\stoblinred(\alpha.V)~W\toblinred$
      $\alpha.(V)~W$,
      and also
      $\alpha.(V)~W\stoblinred\alpha.(V)~W$.

    \item $(M)~(\alpha.V)\toolin\alpha.(M)~V$.  
      \begin{itemize}
	\item Let $0$ be the normal form of $M$. Hence,
	  $(M)~(\alpha.V)\stoblinred(0)~(\alpha.V)\toblinred 0$
	  and
	  $\alpha.(M)~V\stoblinred\alpha.(0)~V\toblinred\alpha.0\toblinred 0$.
	\item Let $\sum\limits_{i=1}^n\alpha_i.B_i$ be the normal form of $M$. Then 
	  $(M)~(\alpha.V)$
	  $\stoblinred(\sum\limits_{i=1}^n\alpha_i.B_i)~(\alpha.V)$
	  $\stoblinred\sum\limits_{i=1}^n\alpha_i.(B_i)~(\alpha.V)$
	  $\stoblinred\sum\limits_{i=1}^n\alpha_i\alpha(B_i)~V$
	  and
	  $\alpha.(M)~V$
	  $\stoblinred\alpha.(\sum\limits_{i=1}^n\alpha_i.B_i)~V$
	  $\stoblinred\alpha.(\sum\limits_{i=1}^n\alpha_i.(B_i)~V)$
	  $\stoblinred\sum\limits_{i=1}^n\alpha_i\alpha(B_i)~V$.
      \end{itemize}

    \item $(0)~M\toolin 0$. Let $V$ be the normal form of $M$. Then
      $(0)~M\stoblinred (0)~V\toblinred 0$.

    \item $(M)~0\toolin 0$.
      \begin{itemize}
	\item Let $0$ be the normal form of $M$, then
	  $(M)~0\stoblinred(0)~0\toblinred 0$.
	\item Let $\sum\limits_{i=1}^n\alpha_i.(B_i)$ be the normal form of $M$. Then
	  $(M)~0$
	  $\stoblinred(\sum\limits_{i=1}^n\alpha_i.(B_i))~0$
	  $\stoblinred\sum\limits_{i=1}^n\alpha_i.(B_i)~0$
	  $\stoblinred\sum\limits_{i=1}^n\alpha_i.0$
	  $\stoblinred\sum\limits_{i=1}^n 0$
	  $\stoblinred 0$.
      \end{itemize}

    \item $(M)~N\toolin(M)~N'$, with $N\toolin N'$. Let $V$ be the normal form of $M$. Then
      $(M)~N$
      $\stoblinred (V)~N\toblinred (V)~N'$
      and
      $(M)~N'\stoblinred (V)~N'$.
      \qedhere
  \end{myenumerate}
\end{proof}

\subsection{Proof of Theorem~\ref{thm:eqred}}\label{proof:eqred}~
\mynobreakpar
\recap{Theorem}{thm:eqred}
For any term $M$ in a confluent fragment of $\llinred$, if $M\stoblineq V$, then $M\stoblinred V'$, with $V\stolineq V'$.
\begin{proof}
  First note that a value can only reduce to another value. This
  follows by direct inspection of the rewriting rules.
  We proceed by induction on the length of the reduction.
  \begin{itemize}
    \item If $M \stoblineq M$, then choose $V'=M$ and note that
      $M\stoblinred M$.
    \item Assume the result true for $M \stoblineq V$: there is a value
      $V'$ such that $M\stoblinred V'$ and $V\stolineq V'$. Let
      $N\toblineq M$. Case distinction:
      \begin{itemize}
	\item $N\tobv M$, then $N\tobv M\stoblinred V'$ which implies
	  $N\stoblinred V'$.
	\item $N\tolineq M$, then either $N\tolinred M$, and then this
	  case is analogous to the previous one, or $M\tolinred N$. Due to
	  the confluence of the subset, there exists a term $L$ such that
	  $N\stoblinred L$ and $V'\stolinred L$, implying that $L$ is a
	  value, thus $V'\stolineq L$. Then we have $V'\stolineq L$ and
	  $V\stolineq V'$, so $L\stolineq V$, closing the case.\qedhere
      \end{itemize}
  \end{itemize}
\end{proof}

\subsection{Proof of Lemma~\ref{lem:substitution2}}\label{proof:substitution2}~

\recap{Lemma}{lem:substitution2}
$\wt{M[x:=B]}=\wt{M}[x:=\Psi(B)]$ with $B$ a base term.
\begin{proof}
  Structural induction on $M$.
  \begin{itemize}
    \item $M=x$. Cases:
      \begin{itemize}
	\item $B=y$. Then $M[x:=B]=y$, and so $\wt{M[x:=B]}=\cont{y}=\cont{x}[y/x]$ and this is equal to $\wt{M}[x:=\Psi(B)]$.
	\item $B=\lambda y\,N$. Then $\wt{M[x:=B]}=\cont{\lambda y\,\wt{N}}=\cont{x}[\lambda y\,\wt{N}/x]=\wt{M}[x:=\Psi(B)]$.
      \end{itemize}
    \item $M=y$. Then $\wt{M[x:=B]}=\wt{M}[x:=\Psi(B)]=\wt{M}$.
    \item $M=0$. Analogous to previous case.
    \item $M=\lambda y\,N$. Then 
      $\wt{(\lambda y\,N)[x:=B]}$
      $=\wt{\lambda y\,(N[x:=B])}$
      $=\cont{\lambda y\,\wt{N[x:=B]}}$, 
      which by the induction hypothesis is 
      $\cont{\lambda y\,\wt{N}[x:=\Psi(B)]}$
      $=(\cont{\lambda y\,\wt{N}})[x:=\Psi(B)]$
      $=\wt{M}[x:=\Psi(B)]$.
    \item $M=(N_1)\ N_2$. Then
      $\wt{M[x:=B]}$
      \!$=\!\wt{((N_1)~N_2)[x:=b]}$
      \!$=\!\wt{(N_1[x:=B])~N_2[x:=B]}$
      $=\tappfgh{\wt{N_1[x:=B]}}{\wt{N_2[x:=B]}}$,
      which, by the induction hypothesis, is equal to
      $\tappfgh{\wt{N_1}[x:=\Psi(B)]}{\wt{N_2}[x:=\Psi(B)]}$,
      which can be rewritten as
      $\tappfgh{\wt{N_1}}{\wt{N_2}}[x:=\Psi(B)]$
      $=\wt{(N_1)~N_2}[x:=\Psi(B)]$
      $=\wt{M}[x:=\Psi(B)]$.
    \item $M=\alpha.N$. Then 
      $\wt{M[x:=B]}$
      $=\wt{(\alpha.N)[x:=B]}$
      $=\wt{\alpha.(N[x:=B])}$
      which is equal to
      $\uncont{f}{\alpha.\wt{N[x:=B]}}$,
      and this, by the induction hypothesis is
      $\uncont{f}{\alpha.\wt{N}[x:=\Psi(B)]}$
      $=(\uncont{f}{\alpha.\wt{N}})[x:=\Psi(B)]$
      $=\wt{\alpha.N}[x:=\Psi(B)]$
      $=\wt{M}[x:=\Psi(B)]$.
    \item $M=N_1+N_2$. Then 
      $\wt{M[x:=\!B]}$
      \!$=\!\wt{(N_1+N_2)[x:=\!B]}$
      \!$=\!\wt{N_1[x:=B]+N_2[x:=B]}$
      $=\uncont{f}{(\wt{N_1[x:=B]}+\wt{N_2[x:=B]})}$,
      which, by the induction hypothesis, is equal to
      $\uncont{f}{(\wt{N_1}[x:=\Psi(B)]+\wt{N_2}[x:=\Psi(B)])}$
      \!$=(\uncont{f}{(\wt{N_1}+\wt{N_2})})[x:=\Psi(B)]$
      $=\wt{N_1+N_2}[x:=\Psi(B)]$
      $=\wt{M}[x:=\Psi(B)]$.
      \qedhere
  \end{itemize}
\end{proof}

\subsection{Proof of Lemma~\ref{lem:lemma2}}\label{proof:lemma2}~

\recap{Lemma}{lem:lemma2}
If $K$ is a base term, then for any term $M$, $(\wt{M})~K\stobalgred M:K$.
\begin{proof}
  Structural induction on $M$.
  \begin{itemize}
    \item $M=\lambda x\,N$. Then $(\wt{\lambda x\,N})~K=(\cont{\lambda x\,\wt{N}})~K$ and by definition of $\Psi$ this is equal to $(\cont{\Psi(\lambda x\,N)})~K \tobalgred  (K) ~\Psi(\lambda x\,N)= \lambda x\,N : K$.
    \item $M=0$. Then $(\wt{0})~K = (0)~K \tobalgred 0 = 0:K$.
    \item $M = M'+N$. Then $(\wt{M'+N})~K = (\lambda f\, (\wt{M'}+\wt{N})~f)~K$ which $\tobalgred$-reduces to $(\wt{M'}+\wt{N})~K \tobalgred  (\wt{M'})~K+(\wt{N})~K $  which $\tobalgred$-reduces by the induction hypothesis to $M':K+N:K = M'+N:K$.
    \item $M = \alpha.N$. Then $(\wt{\alpha.N})~K = (\lambda f\, (\alpha. \wt{N})~f)~K \tobalgred (\alpha. \wt{N})~K$ which $\tobalgred$-reduces to $\alpha.(\wt{N})~K)$ and this, by the induction hypothesis, $\tobalgred$-reduces to $ \alpha.(N:K) = \alpha.N : K$. 
    \item $M=(M')~N$. Then $(\wt{(M')~N})~K=(\tappfgh{\wt{M'}}{\wt N})~K$ which $\tobalgred$-reduces to $(\wt{M'})~\lambda g\,(\wt{N})~\lambda h\,((g)~h)~K$. 
      Since $\lambda g\,(\wt{N})~\lambda h\,((g)~h)~K$ is a value, by the induction hypothesis 
      $(\wt{M'})~\lambda g\,(\wt{N})~\lambda h\,((g)~h)~K$
      reduces to $M':\lambda g\,(\wt{N})~\lambda h\,((g)~h)~K$.
      We do a second induction, over $M'$, to prove that $M':\lambda g\,(\wt{N})~\lambda h\,((g)~h)~K\tobalgred (M')~N:K$.
      \begin{itemize} 
	\item If $M'=(M_1)~M_2$, then $M':\lambda g\,(\wt{N})~\lambda h\,((g)~h)~K=((M_1)~M_2)~N:K=(M')~N:K$.
	\item If $M'$ is a base term, then $M':\lambda g\,(\wt{N})~\lambda h\,((g)~h)~K
	  $ is by definition equal to the term $
	  (\lambda g\,(\wt{N})~\lambda h\,((g)~h)~K)~\Psi(M')\tobalgred(\wt{N})~\lambda h\,((\Psi(M'))~h)~K$  which by the main induction hypothesis $\tobalgred$-reduces to  $ N : \lambda h\,((\Psi(M'))~h)~K$, and this is equal to $(M')~N:K$.
	\item If $M' = \alpha.M_1$, then the term $M':\lambda g\,(\wt{N})~\lambda h\,((g)~h)~K$ is equal to $\alpha.M_1:\break\lambda g\,(\wt{N})~\lambda h\,((g)~h)~K = \alpha.(M_1:\lambda g\,(\wt{N})~\lambda h\,((g)~h)~K)$ which by the second induction hypothesis $\tobalgred$-reduces to $\alpha.((M_1)~N:K)=(\alpha.M_1)~N:K = (M')~N : K$.
	\item If $M' = M_1+M_2$, then $M':\lambda g\,(\wt{N})~\lambda h\,((g)~h)~K$ is equal to the term $M_1+M_2 :\lambda g\,(\wt{N})~\lambda h\,((g)~h)~K$ which is equal to $M_1 :\lambda g\,(\wt{N})~\lambda h\,((g)~h)~K+M_2 :\lambda g\,(\wt{N})~\lambda h\,((g)~h)~K$ which $\tobalgred$-reduces by the second induction hypothesis to $(M_1)~N:K+(M_2)~N : K = (M_1+M_2)~N:K = (M')~N : K$.
	\item If $M'=0$ then $M:\lambda g\,(\wt{N})~\lambda h\,((g)~h)~K= 0 :\lambda g\,(\wt{N})~\lambda h\,((g)~h)~K = 0= (0)~N:K = (M')~N : K$.\qedhere
      \end{itemize}
  \end{itemize}
\end{proof}

\subsection{Proof of Lemma~\ref{lem:lemma3alg}}\label{proof:lemma3alg}~

\recap{Lemma}{lem:lemma3alg}
If $M\tolinred N$ then for all $K$ base term, $M:K\stoalgred N:K$.
\begin{proof}
  Case by case on the rules $\tolinred$.
  \begin{description}
    \item[Rules $A_r$]~
      \begin{itemize}
	\item $(B)~(M+N)\tolinred (B)~M+(B)~N$, with $B$ being a base term. Then $(B)~(M+N):K= M+N :\lambda f\,((\Psi(B))~f)~K = M :\lambda f\,((\Psi(B))~f)~K  + N:\lambda f\,((\Psi(B))~f)~K  = (B)~M : K + (B)~N : K = (B)~M+  (B)~N : K$.
	\item $(B)~\alpha.M\tolinred\alpha.(B)~M$, with $B$ a base
	  term. Then $(B)~\alpha.M:K$ is equal to the term
	  $\alpha.M:\lambda f\,((\Psi(B))~f)~K$ $= \alpha.(M : \lambda
	  f\,((\Psi(B))~f)~K) = \alpha.((B)~M:K)$, which is $\alpha.(B)~M : K$.
	\item $(B)~0\tolinred 0$, with $B$ a base term. Then $(B)~0:K=0 :\lambda f\,((\Psi(B))~f)~K = 0 =0:K$.
      \end{itemize}
    \item[Rules $A_l$]~\mynobreakpar
      \begin{itemize}
	\item $(M+N)~V\tolinred (M)~V+(N)~V$, with $V$ being a value. Then $(M+N)~V : K= (M)~V+(N)~V : K$.
	\item $(\alpha.M)~V\tolinred\alpha.(M)~V)$, with $V$ being a value. Then $(\alpha.M)~V : K = \alpha.(M)~V : K$.
	\item $(0)~V\tolinred 0$, with $V$ a value. Then $(0)~V : K = 0 = 0:K$.
      \end{itemize}
    \item[Rules $F$ and $S$]~
      \begin{itemize}
	\item $\alpha.(M+N)\tolinred \alpha.M+\alpha.N$. Then $\alpha.(M+N):K=\alpha.(M:K+N:K)\toalgred\alpha.(M:K)+\alpha.(N:K)=\alpha.M+\alpha.N:K$.
	\item $\alpha.M+\beta.M\tolinred(\alpha+\beta).M$. Then $\alpha.M+\beta.M:K=\alpha.(M:K)+\beta.(M:K)\toalgred(\alpha+\beta).(M:K)=(\alpha+\beta).M:K$.
	\item $\alpha.M+M\tolinred(\alpha+1).M$. Then  $\alpha.M+M:K=\alpha.M:K+M:K=\alpha.(M:K)+M:K\toalgred(\alpha+1).(M:K)=(\alpha+1).M:K$.
	\item $M+M\tolinred(1+1).M$. Then  $M+M:K=M:K+M:K\toalgred(1+1).(M:K)=(1+1).M:K$.
	\item $0+M\tolinred M$. Then  $0+M:K=(0:K)+(M:K)=0+(M:K)\toalgred M:K$.
	\item $\alpha.(\beta.M)\tolinred(\alpha\beta).M$. Then  $\alpha.(\beta.M):K=\alpha.(\beta.M:K)=\alpha.(\beta.(M:K))$ which $\toalgred$-reduces to $(\alpha\beta).(M:K)=(\alpha\beta).M:K$.
	\item $1.M\tolinred M$. Then $1.M:K=1.(M:K)\toalgred M:K$.
	\item $0.M\tolinred 0$. Then  $0.M:K=0.(M:K)\toalgred 0=0:K$.
	\item $\alpha.0\tolinred 0$. Then  $\alpha.0:K=\alpha.(0:K)=\alpha.0\toalgred 0=0:K$.
      \end{itemize}
    \item[Rules $Asso$ and $Com$]~
      \begin{itemize}
	\item $M+(N+L)\tolinred (M+N)+L$. Then $M+(N+L):K=M:K+(N+L:K)=M:K+(N:K+L:K)\toalgred (M:K+N:K)+L:K=M+N:K+L:K=(M+N)+L:K$.
	\item $M+N\tolinred N+M$. Then $M+N:K=M:K+N:K\toalgred N:K+M:K=N+M:K$.
      \end{itemize}
    \item[Rules $\xi$ and $\xi_{\lalin}$] Assume $M\tolinred M'$, and assume that for all $K$ base term, $M:K\stoalgred M':K$. We show that the result also holds for each contextual rule.
      \begin{itemize}
	\item $M+N\tolinred M'+N$. Then $M+N:K=M:K+N:K\stoalgred M':K+N:K=M'+N:K$.
	\item $N+M\tolinred N+M'$, analogous to previous case.
	\item $\alpha.M\tolinred\alpha.M'$. Then $\alpha.M:K=\alpha.(M:K)\stoalgred \alpha.(M':K)=\alpha.M':K$.
	\item $(V)~M\tolinred (V)~M'$. Case by case:
	  \begin{itemize}
	    \item $V=B$.  Then $(B)~M:K = M: \lambda f\,((\Psi(B))~f)~K$ which $\toalgred$-reduces by the induction hypothesis to $M': \lambda f\,((\Psi(B))~f)~K =(B)~M': K$.
	    \item $V=0$. Then $(0)~M:K = 0 = (0)~M':K$.
	    \item $V=\alpha.W$. Then $(\alpha.W)~M  : K = \alpha.(W)~M : K=\alpha.((W)~M :K)$  which $\toalgred $-reduces by the induction hypothesis to $\alpha.((W)~M':K) = \alpha.(W)~M':K=(\alpha.W)~M':K$.
	    \item $V=V_1+V_2$. Then $(V_1+V_2)~M  : K = (V_1)~M  + (V_2)~M: K= (V_1)~M : K + (V_2)~M: K$  which $\toalgred $-reduces by the induction hypothesis to $(V_1)~M' : K + (V_2)~M': K= (V_1)~M'+(V_2)~M':K=(V_1+V_2)~M' : K$.
	  \end{itemize}
	\item $(M)~N\tolinred (M')~N$ Case by case:
	  \begin{itemize}
	    \item $M=B$. Absurd since a base term cannot reduce.
	    \item $M=\alpha.M_1$. Case by case on the possible $\tolinred$-reductions of $M$:
	      \begin{itemize}
		\item $M'=\alpha.M_1'$ with $M_1\tolinred M_1'$. Then $(\alpha.M_1)~N:K=\alpha.(M_1)~N:K=\alpha.((M_1)~N:K)$ which by the induction hypothesis $\toalgred$-reduces to $\alpha.((M_1')~N:K)=\alpha.(M_1')~N:K=(\alpha.M_1')~N:K$.
		\item $M=\alpha.(\beta.M_3)$ and $M'=(\alpha\beta).M_3$. Then $(\alpha.(\beta.M_3))~N:K=\alpha.(\beta.((M_3)~N:K))\toalgred(\alpha\beta).((M_3)~N:K)=((\alpha\beta).M_3)~N:K$.
		\item $M=\alpha.(L_1+L_2)$ and $M'=\alpha.L_1+\alpha.L_2$. Then $(\alpha.(L_1+L2))~N:K=\alpha.((L_1)~N:K+(L_2)~N:K)\toalgred\alpha.((L_1)~N:K)+\alpha.((L_2)~N:K)=(\alpha.L_1+\alpha.L2)~N:K$.
		\item $\alpha=1$ and $M'=M_1$. Then $(1.M_1)~N:K=1.((M_1)~N:K)$ which \toalgred-reduces to $(M_1)~N:K$.
		\item $\alpha=0$ and $M'=0$. Then $(0.M_1)~N:K=0.((M_1)~N:K)\toalgred 0=(0)~N:K$.
		\item $M_1=0$ and $M'=0$. Then $(\alpha.0)~N:K=\alpha.((0)~N:K)=\alpha.0\toalgred 0=(0)~N:K$.
	      \end{itemize}
	    \item $M=M_1+M_2$. Case by case on the possible $\tolinred$-reductions of $M$:
	      \begin{itemize}
		\item $M'=M_1'+M_2$ with $M_1\tolinred M'_1$. Then $(M_1+M_2)~N:K=(M_1)~N:K+(M_2)~N:K$ which by the induction hypothesis $\toalgred$-reduces to $(M_1')~N:K+(M_2)~N:K=(M_1'+M_2)~N:K$.
		\item $M'=M_1+M_2'$ with $M_2\tolinred M'_2$. Analogous to previous case.
		\item $M_2=L_1+L_2$ and $M'=(M_1+L_1)+L_2$. Then $(M_1+(L_1+L_2))~N:K=(M_1)~N:K+((L_1)~N:K+(L_2)~N:K)$ and this $\toalgred$-reduces to $((M_1)~N:K+(L_1)~N:K)+(L_2)~N:K=((M_1+L_1)+L_2)~N:K$.
		\item $M_1=L_1+L_2$ and $M'=L_1+(L_2+M_2)$. Analogous to previous case.
		\item $M'=M_2+M_1$. Then $(M_1+M_2)~N:K=(M_1)~N:K+(M_2)~N:K\toalgred (M_2)~N:K+(M_1)~N:K=(M_2+M_1)~N:K$.
		\item $M_1=\alpha.M_3$, $M_2=\beta.M_3$ and $M'=(\alpha+\beta).M_3$. Then $(\alpha.M_3+\beta.M_3)~N:K=\alpha.((M_3)~N:K)+\beta.((M_3)~N:K)$ which \toalgred-reduces to $(\alpha+\beta).((M_3)~N:K)=((\alpha+\beta).M_3)~N:K$.
		\item $M_1=\alpha.M_3$, $M_2=M_3$ and $M'=(\alpha+1).M_3$. Analogous to previous case.
		\item $M_1=M_2$ and $M'=(1+1).M_1$. Analogous to previous case.
	      \end{itemize}
	    \item $M=0$. Absurd since $0$ does not reduce.
	    \item $M=(M_1)~M_2$. Then the term $((M_1)~M_2)~N:K$ is equal to $(M_1)~M_2:\lambda g\,(\wt N)~\lambda h\,((g)~h)~K$, which by the induction hypothesis $\toalgred$-reduces to $M':\lambda g\,(\wt N)~\lambda h\,((g)~h)~K$.
	      We do a second induction, over $M'$, to prove that $M':\lambda g\,(\wt{N})~\lambda h\,((g)~h)~K\toalgred (M')~N:K$.
	      \begin{itemize} 
		\item If $M'=(M'_1)~M'_2$, then $M':\lambda g\,(\wt{N})~\lambda h\,((g)~h)~K$ is equal to $((M'_1)~M'_2)~N:K=(M')~N:K$.
		\item $M'$ cannot be a base term since from $(M_1)~M_2$ it is not possible to arrive to a base term using only $\tolinred$.
		\item If $M' = \alpha.M'_1$, then $M':\lambda g\,(\wt{N})~\lambda h\,((g)~h)~K$ is equal to the term $\alpha.M'_1:\lambda g\,(\wt{N})~\lambda h\,((g)~h)~K=\alpha.(M'_1:\lambda g\,(\wt{N})~\lambda h\,((g)~h)~K)$ which $\toalgred$-reduces by the induction hypothesis to $\alpha.((M'_1)~N:K) = (\alpha.M'_1)~N:K = (M')~N : K$.
		\item If $M' = M'_1+M'_2$, then the term $M':\lambda g\,(\wt{N})~\lambda h\,((g)~h)~K$ is equal to $M'_1+M'_2 :\lambda g\,(\wt{N})~\lambda h\,((g)~h)~K$ which is equal to $M'_1 :\lambda g\,(\wt{N})~\lambda h\,((g)~h)~K+M'_2 :\lambda g\,(\wt{N})~\lambda h\,((g)~h)~K$ which $\toalgred$-reduces by the induction hypothesis to $(M'_1)~N:K+(M'_2)~N : K = (M'_1+M'_2)~N:K = (M')~N : K$.
		\item If $M'=0$ then $M':\lambda g\,(\wt{N})~\lambda h\,((g)~h)~K$ is equal $0 :\lambda g\,(\wt{N})~\lambda h\,((g)~h)~K$ and this to $0= (0)~N:K = (M')~N : K$.\qedhere
	      \end{itemize}
	  \end{itemize}
      \end{itemize}
  \end{description} 
\end{proof}

\subsection{Proof of Lemma~\ref{lem:lemma3}}\label{proof:lemma3}~

\recap{Lemma}{lem:lemma3}
If $M\toblinred N$ then for all $K$ base term, $M:K\stobalgred N:K$.
\begin{proof}
  Case by case on the rules $\toblinred$.
  \begin{description}
    \item[Rule $\beta_v$]
      $(\lambda x\ M)\ B :K$
      $=B : \lambda f\,((\Psi(\lambda x\ M))~f)~K$
      $=(\lambda f\,((\Psi(\lambda x\ M))~f)~K)~\Psi(B)$
      $\tobn((\Psi(\lambda x\,M))\,\Psi(B))\,K$
      $=((\lambda x\,\wt{M})\,\Psi(B))\,K$
      $\tobn\wt{M}[x:=\Psi(B)]\,K$,
      which, by Lemma~\ref{lem:substitution2}, is equal to
      $\wt{M[x:=B]}\,K$, and this, by Lemma~\ref{lem:lemma2},
      \stobalgred-reduces to
      $M[x:=B]:K$.
    \item[Algebraic rules] If $M\tolinred N$, then by Lemma~\ref{lem:lemma3alg} $M:K\stoalgred N:K$ which implies that $M:K\stobalgred N:K$.
    \item[Rules $\xi$ and $\xi_{\lalin}$] If $M\tolinred M'$, then we use Lemma~\ref{lem:lemma3alg} to close the case. 
      Assume $M\tobv M'$, and assume that for all $K$ base term, $M:K\stobalgred M':K$. We show that the result also holds for each contextual rule.
      \begin{itemize}
	\item $M+N\tobv M'+N$. Then $M+N:K=M:K+N:K\stobalgred M':K+N:K=M'+N:K$.
	\item $N+M\tobv N+M'$, analogous to previous case.
	\item $\alpha.M\tobv\alpha.M'$. Then $\alpha.M:K=\alpha.(M:K)\stobalgred \alpha.(M':K)=\alpha.M':K$.
	\item $(V)~M\tobv (V)~M'$. Case by case:
	  \begin{itemize}
	    \item $V=B$.  Then $(B)~M:K = M: \lambda f\,((\Psi(B))~f)~K$ which $\tobalgred $-reduces by the induction hypothesis to $M': \lambda f\,((\Psi(B))~f)~K =(B)~M': K$.
	    \item $V=0$. Then $(0)~M:K = 0 = (0)~M':K$.
	    \item $V=\alpha.W$. Then $(\alpha.W)~M  : K = \alpha.(W)~M : K=\alpha.((W)~M :K)$  which $\tobalgred $-reduces by the induction hypothesis to $\alpha.((W)~M':K) = \alpha.(W)~M':K=(\alpha.W)~M':K$.
	    \item $V=V_1+V_2$. Then $(V_1+V_2)~M  : K = (V_1)~M  + (V_2)~M: K= (V_1)~M : K + (V_2)~M: K$  which $\tobalgred $-reduces by the induction hypothesis to $(V_1)~M' : K + (V_2)~M': K= (V_1)~M'+(V_2)~M':K=(V_1+V_2)~M' : K$.
	  \end{itemize}
	\item $(M)~N\tobv (M')~N$ Case by case:
	  \begin{itemize}
	    \item $M=B$. Absurd since a base term cannot reduce.
	    \item $M=\alpha.M_1$. The only possible $\tobv$-reduction from $M$ is $M'=\alpha.M_1'$ with $M_1\tobv M_1'$. Then $(\alpha.M_1)~N:K=\alpha.(M_1)~N:K=\alpha.((M_1)~N:K)$ which by the induction hypothesis $\tobalgred$-reduces to $\alpha.((M_1')~N:K)=\alpha.(M_1')~N:K=(\alpha.M_1')~N:K$.
	    \item $M=M_1+M_2$. Case by case on the possible $\tobv$-reductions of $M$:
	      \begin{itemize}
		\item $M'=M_1'+M_2$ with $M_1\tobv M'_1$. Then $(M_1+M_2)~N:K=(M_1)~N:K+(M_2)~N:K$ which by the induction hypothesis $\tobalgred$-reduces to $(M_1')~N:K+(M_2)~N:K=(M_1'+M_2)~N:K$.
		\item $M'=M_1+M_2'$ with $M_2\tobv M'_2$. Analogous to previous case.
	      \end{itemize}
	    \item $M=0$. Absurd since $0$ does not reduce.
	    \item $M=(M_1)~M_2$. Then the term $((M_1)~M_2)~N:K$ is equal to $(M_1)~M_2:\lambda g\,(\wt N)~\lambda h\,((g)~h)~K$, which $\tobalgred$-reduces, by the induction hypothesis, to $M':\lambda g\,(\wt N)~\lambda h\,((g)~h)~K$.
	      We do a second induction, over $M'$, to prove that $M':\lambda g\,(\wt{N})~\lambda h\,((g)~h)~K$ $\tobalgred$-reduces to $(M')~N:K$.
	      \begin{itemize} 
		\item If $M'=(M'_1)~M'_2$, then $M':\lambda g\,(\wt{N})~\lambda h\,((g)~h)~K$ is equal to $((M'_1)~M'_2)~N:K=(M')~N:K$.
		\item If $M'$ is a base term, then the term $M':\lambda g\,(\wt{N})~\lambda h\,((g)~h)~K$ is equal to $(\lambda g\,(\wt{N})~\lambda h\,((g)~h)~K)~\Psi(M')$ which $\tobalgred$-reduces to $(\wt{N})~\lambda h\,((\Psi(M'))~h)~K$  which, by Lemma~\ref{lem:lemma2}, $\tobalgred$-reduces $ N : \lambda h\,((\Psi(M'))~h)~K = (M')~N:K$.
		\item If $M' = \alpha.M'_1$, then $M':\lambda g\,(\wt{N})~\lambda h\,((g)~h)~K$ is equal to the term $\alpha.M'_1:\lambda g\,(\wt{N})~\lambda h\,((g)~h)~K=\alpha.(M'_1:\lambda g\,(\wt{N})~\lambda h\,((g)~h)~K)$ which $\tobalgred$-reduces by the induction hypothesis to $\alpha.((M'_1)~N:K) = (\alpha.M'_1)~N:K = (M')~N : K$.
		\item If $M' = M'_1+M'_2$, then the term $M':\lambda g\,(\wt{N})~\lambda h\,((g)~h)~K$ is equal to $M'_1+M'_2 :\lambda g\,(\wt{N})~\lambda h\,((g)~h)~K$ which is equal to $M'_1 :\lambda g\,(\wt{N})~\lambda h\,((g)~h)~K+M'_2 :\lambda g\,(\wt{N})~\lambda h\,((g)~h)~K$ which $\tobalgred$-reduces by the induction hypothesis to $(M'_1)~N:K+(M'_2)~N : K = (M'_1+M'_2)~N:K = (M')~N : K$.
		\item If $M'=0$ then $M':\lambda g\,(\wt{N})~\lambda h\,((g)~h)~K$ is equal to $0 :\lambda g\,(\wt{N})~\lambda h\,((g)~h)~K = 0= (0)~N:K = (M')~N : K$.\qedhere
	      \end{itemize}
	  \end{itemize}
      \end{itemize}
  \end{description} 
\end{proof}

\subsection{Proof of Lemma~\ref{lem:inverse-term}}\label{proof:inverse-term}~

\recap{Lemma}{lem:inverse-term} For any term $M$, $\overline{(\wt M)~k}=M$.

\begin{proof}
  By induction on $M$.
  \begin{itemize}
    \item Case $x$. Then $\sigma(\wt x)=\sigma(\lambda k\,(k)~x)=\overline{(k)~x}=\underline{k}[\psi(x)]=\psi(x)=x$.
    \item Case $\lambda x\,M$. Then 
      $\sigma(\wt{\lambda x\,M})$
      $=\sigma(\lambda k\,(k)~\lambda x\,\wt M)$
      $=\overline{k(\lambda x\,\wt M)}$
      $=\underline{k}[\psi(\lambda x\,\wt M)]$
      $=\psi(\lambda x\,\wt M)$
      $=\lambda x\,\sigma(\wt M)$,
      which, by the induction hypothesis, is equal to
      $=\lambda x\,M$.

    \item Case $MN$. Then 
      $\sigma(\wt{(M)~N})$
      $=\sigma(\lambda k\,(\wt M)~\lambda b_{1}\,(\wt N)~\lambda b_{2}\,((b_{1})~b_{2})~k)$
      which is equal to
      $\overline{(\wt M)~\lambda b_{1}\,(\wt N)~\lambda b_{2}\,((b_{1})~b_{2})~k}$
      $=\underline{\lambda b_{1}\,(\wt N)~\lambda b_{2}\,((b_{1})~b_{2})~k}[\sigma(\wt M)]$
      which is equal to
      $\underline{k}[(\sigma(\wt M))~\sigma(\wt N)]$
      $=(\sigma(\wt M))~\sigma(\wt N)$,
      which, by the induction hypothesis, is equal to
      $=(M)~N$.

    \item Case $0$. Then $\sigma(\wt 0)=\sigma(0)=0$
    \item Case $\alpha.M$. Then 
      $\sigma(\wt{\alpha.M})$
      $=\sigma(\lambda k\,(\alpha.\wt M)~k)$
      $=\overline{(\alpha.\wt M)~k}$
      $=\underline{k}[\sigma(\alpha.\wt M)]$
      $=\sigma(\alpha.\wt M)$
      $=\alpha.\sigma(\wt M)$,
      which, by the induction hypothesis, is equal to
      $=\alpha.M$.

    \item Case $M+N$. Then
      $\sigma(\wt{M+N})$
      $=\sigma(\lambda k\,(\wt M+\wt N)~k)$
      $=\overline{(\wt M+\wt N)~k}$
      $=\underline{k}[\sigma(\wt M+\wt N)]$
      $=\sigma(\wt M+\wt N)$
      $=\sigma(\wt M)+\sigma(\wt N)$,
      which, by the induction hypothesis, is equal to
      $=M+N$.
      \qedhere
  \end{itemize}
\end{proof}

\subsection{Proof of Lemma \ref{lem:inverse-value}}\label{proof:inverse-value}~

\recap{Lemma}{lem:inverse-value} For any value $V$, $\overline{V:k}=V$.
\begin{proof}
  By induction on $V$.
  \begin{itemize}
    \item Case $0$. Then $\overline{0:k}=\overline{0}=0$.
    \item Case $B$. Then $\overline{B:k}=\overline{(k)~\Psi(B)}=\sigma(\lambda k\,(k)~\Psi(B))=\sigma(\wt B)=B$
      by Lemma \ref{lem:inverse-term}. 
    \item Case $\alpha.V$. Then $\overline{\alpha.V:k}=\overline{\alpha.(V:k)}=\alpha.\overline{V:k}=\alpha V$
      by the induction hypothesis.
    \item Case $U+V$. Then $\overline{U+V:k}=\overline{U:k+V:k}=\overline{U:k}+\overline{V:k}=U+V$
      by the induction hypothesis.
      \qedhere
  \end{itemize}
\end{proof}

\subsection{Proof of Lemma~\ref{lem:inverse-step}}\label{proof:inverse-step}~

\recap{Lemma}{lem:inverse-step} For any computation $D$, if $D\tobalgred D'$
then $\overline{D}\stoblinred\overline{D'}$. Also, if $D\toalgred D'$
then $\overline{D}\stolinred\overline{D'}$.

\bigskip
To prove this lemma, we need several intermediary results.

\begin{lemma}\label{lem:substitution-lemma}
  The following equalities hold.
  \begin{myenumerate}
    \item $\psi(B_{1})[x:=\psi(B)]=\psi(B_{1}[x:=B])$
    \item $\sigma(T)[x:=\psi(B)]=\sigma(T[x:=B])$
    \item $\overline{C}[x:=\psi(B)]=\overline{C[x:=B]}$
    \item $\underline{K}[M][x:=\psi(B)]=\underline{K[x:=B]}[M[x:=\psi(B)]]$
  \end{myenumerate}
\end{lemma}
\begin{proof}
  We prove simultaneously the four properties by induction on the structure of $B_{1}$, $T$, $C$, and $K$.
  \begin{myenumerate}
    \item Cases for $B_{1}$.
      \begin{itemize}
	\item Case $x$. Then $\psi(x)[x:=\psi(B)]=x[x:=\psi(B)]=\psi(B)=\psi(x[x:=B])$.
	\item Case $y\neq x$. Then $\psi(y)[x:=\psi(B)]=y[x:=\psi(B)]=y=\psi(y)=\psi(y[x:=B])$.
	\item Case $\lambda y\,S$. Then 
	  $\psi(\lambda x\,S)[x:=\psi(B)]$
	  $=(\lambda y\,\sigma(S))[x:=\psi(B)]$,
	  which, by the induction hypothesis, is equal to
	  $\lambda y\,\sigma(S[x:=B])$
	  $=\psi((\lambda y\,S)[x:=B])$.
      \end{itemize}
    \item Cases for $T$.
      \begin{itemize}
	\item Case $\lambda k\,C$. Then 
	  $\sigma(\lambda k\,C)[x:=\psi(B)]$
	  $=\overline{C}[x:=\psi(B)]$,
	  which, by the induction hypothesis, is equal to
	  $\overline{C[x:=B]}$
	  $=\sigma((\lambda k\,C)[x:=B])$.

	\item Case $0$. Then $\sigma(0)[x:=\psi(B)]=0=\sigma(0[x:=B])$.
	\item Case $\alpha.T$. Then 
	  $\sigma(\alpha.T)[x:=\psi(B)]$
	  $=(\alpha.\sigma(T))[x:=\psi(B)]$,
	  which, by the induction hypothesis, is equal to
	  $\alpha.\sigma(T[x:=B])$
	  $=\sigma((\alpha.T)[x:=B])$.

	\item Case $T_{1}+T_{2}$. Then 
	  $\sigma(T_{1}+T_{2})[x:=\psi(B)]$
	  $=(\sigma(T_{1})+\sigma(T_{2}))[x:=\psi(B)]$,
	  which, by the induction hypothesis, is equal to
	  $\sigma(T_{1}[x:=B])+\sigma(T_{2}[x:=B])$
	  $=\sigma((T_{1}+T_{2})[x:=B])$.
      \end{itemize}
    \item Cases for $C$.
      \begin{itemize}
	\item Case $(K)~B_{1}$. Then 
	  $\overline{(K)~B_{1}}[x:=\psi(B)]$
	  $=\underline{K}[\psi(B_{1})][x:=\psi(B)]$,
	  which, by the induction hypothesis, is equal to
	  $\underline{K[x:=B]}[\psi(B_{1})[x:=\psi(B)]]$,
	  which, by the induction hypothesis, is 
	  $\underline{K[x:=B]}[\psi(B_{1}[x:=B])]$
	  $=\overline{((K)~B_{1})[x:=B]}$.

	\item Case $((B_{1})~B_{2})~K$. Then
	  $\overline{((B_{1})~B_{2})~K}[x:=\psi(B)]$
	  $=\underline{K}[(\psi(B_{1}))~\psi(B_{2})][x:=\psi(B)]$,
	  which, by the induction hypothesis, is 
	  $\underline{K[x:=B]}[((\psi(B_{1}))~\psi(B_{2}))[x:=\psi(B)]]$,
	  which, by the induction hypothesis, is equal to
	  $\underline{K[x:=B]}[(\psi(B_{1}[x:=B]))~\psi(B_{2}[x:=B])]$
	  $=\overline{(((B_{1})~B_{2})~K)[x:=B]}$.

	\item Case $(T)~K$. Then 
	  $\overline{(T)~K}[x:=\psi(B)]$
	  $=\underline{K}[\sigma(T)][x:=\psi(B)]$,
	  which, by the induction hypothesis, is equal to
	  $\underline{K[x:=B]}[\sigma(T)[x:=\psi(B)]]$,
	  which, by the induction hypothesis, is equal to
	  $\underline{K[x:=B]}[\sigma(T[x:=B])]$
	  $=\overline{((T)~K)[x:=B]}$.
      \end{itemize}
    \item Cases for $K$.
      \begin{itemize}
	\item Case $k$. Then $\underline{k}[M][x:=\psi(B)]=M[x:=\psi(B)]=\underline{k[x:=B]}[M[x:=\psi(B)]]$.
	\item Case $\lambda b\,B_{1}bK$. Then 
	  $\underline{\lambda b\,((B_{1})~b)~K}[M][x:=\psi(B)]$
	  $=\underline{K}[(\psi(B_{1}))~M][x:=\psi(B)]$,
	  which, by the induction hypothesis, is 
	  $\underline{K[x:=B]}[((\psi(B_{1}))~M)[x:=\psi(B)]]$,
	  which, by the induction hypothesis, is equal to
	  $\underline{K[x:=B]}[(\psi(B_{1}[x:=B]))~M[x:=\psi(B)]]$
	  $=\underline{(\lambda b\,((B_{1})~b)~K)[x:=B]}[M[x:=\psi(B)]]$.

	\item Case $\lambda{b_{1}}\,(T)~\lambda b_2\,((b_{1})~b_{2})~K$. Then
	  $\underline{\lambda b_1\,(T)~\lambda b_2\,((b_{1})~b_{2})~K}[M][x:=\psi(B)]$
	  is the term
	  $\underline{K}[(M)~\sigma(T)][x:=\psi(B)]$,
	  which, by the induction hypothesis, is equal to
	  $\underline{K[x:=B]}[((M)~\sigma(T))[x:=\psi(B)]]$,
	  which, by the induction hypothesis, is equal to
	  $\underline{K[x:=B]}[(M[x:=\psi(B)])~\sigma(T[x:=B])]$
	  and this is finally equal to the term
	  $\underline{(\lambda b_1\,(T)~\lambda b_2\,((b_{1})~b_{2})~K)[x:=B]}[M[x:=\psi(B)]]$.
	  \qedhere
      \end{itemize}
  \end{myenumerate}
\end{proof}

\begin{lemma}\label{lem:continuation-composition}
  For all terms $M$ and continuations
  $K_{1}$ and $K_{2}$, 
  $
  \underline{K_{1}}[\underline{K_{2}}[M]]\!=\!\underline{K_{2}[k:=K_{1}]}[M]
  $.
\end{lemma}
\begin{proof}
  By induction on the structure of $K_{2}$. 
  \begin{itemize}
    \item Case $k$. Then $\underline{K_{1}}[\underline{k}[M]]=\underline{K_{1}}[M]=\underline{k[k:=K_{1}]}[M]$.
    \item Case $\lambda b\,((B)~b)~K$. Then 
      $\underline{K_{1}}[\underline{\lambda b\,((B)~b)~K}[M]]$
      $=\underline{K_{1}}[\underline{K}[(\psi(B))~M]]$,
      which, by the induction hypothesis, is 
      $\underline{K[k:=K_{1}]}[(\psi(B))~M]$
      $=\underline{(\lambda b\,((B)~b)~K)[k:=K_{1}]}[M]$.

    \item Case $\lambda b_1\,(T)~\lambda b_2\,((b_{1})~b_{2})~K$. Then 
      $\underline{K_{1}}[\underline{\lambda b_1\,(T)~\lambda b_2\,((b_{1})~b_{2})~K}[M]]$
      which is equal to
      $\underline{K_{1}}[\underline{K}[(M)~\sigma(T)]]$,
      which, by the induction hypothesis, is 
      $\underline{K[k:=K_{1}]}[M\sigma(T)]$, which is equal to the term
      $\underline{(\lambda b_1\,(T)~\lambda b_2\,((b_{1})~b_{2})~K)[k:=K_{1}]}[M]$.
      \qedhere
  \end{itemize}
\end{proof}

\begin{lemma}\label{lem:continuation-substitution}
  For all $K$ and $C$,
  $\underline{K}[\overline{C}]=\overline{C[k:=K]}$.
\end{lemma}
\begin{proof}
  By induction on the structure of $C$, using Lemma \ref{lem:continuation-composition}
  where necessary.
  \begin{itemize}
    \item Case $(K_{2})~B$. Then 
      $\underline{K}[\overline{(K_{2})~B}]$
      $=\underline{K}[\underline{K_{2}}[\psi(B)]]$,
      which, by Lemma \ref{lem:continuation-composition}, is equal to
      $\underline{K_{2}[k:=K]}[\psi(B)]$
      $=\overline{((K_{2})~B)[k:=K]}$.

    \item Case $((B_{1})~B_{2})~K_{2}$. Then 
      $\underline{K}[\overline{((B_{1})~B_{2})~K_{2}}]$
      $=\underline{K}[\underline{K_{2}}[(\psi(B_{1}))~\psi(B_{2})]]$,
      which, by Lemma \ref{lem:continuation-composition}, is equal to
      $\underline{K_{2}[k:=K]}[(\psi(B_{1}))~\psi(B_{2})]$
      $=\overline{(((B_{1})~B_{2})~K_{2})[k:=K]}$.

    \item Case $(T)~K_{2}$. Then 
      $\underline{K}[\overline{(T)~K_{2}}]$
      $=\underline{K}[\underline{K_{2}}[\sigma(T)]]$,
      which, by Lemma \ref{lem:continuation-composition}, is equal to
      $=\underline{K_{2}[k:=K]}[\sigma(T)]$
      $=\overline{((T)~K_{2})[k:=K]}$.
      \qedhere
  \end{itemize}
\end{proof}

\begin{lemma}\label{lem:continuation-step}
  For any continuation $K$ and term
  $M$, if $M\toblinred M'$, then $\underline{K}[M]\toblinred\underline{K}[M']$.
\end{lemma}
\begin{proof}
  By induction on the structure of $K$.
  \begin{itemize}
    \item Case $k$. Then $\underline{k}[M]=M\toblinred M'=\underline{k}[M']$.
    \item Case $\lambda b\,((B)~b)~K$. Then $\psi(B)M\toblinred\psi(B)M'$ since $\psi(B)$
      is a base term, and 
      $\underline{(\lambda b\,((B)~b)~K)}[M]$
      $=\underline{K}[(\psi(B))~M]$,
      which, by the induction hypothesis, \toblinred-reduces to
      $\underline{K}[(\psi(B))~M']$
      $=\underline{\lambda b\,((B)~b)~K}[M']$.

    \item Case $\lambda b_1\,(T)~\lambda b_2\,((b_{1})~b_{2})~K$. Then we have
      $M\sigma(T)\toblinred M'\sigma(T)$. Hence, we have
      $\underline{\lambda b_1\,(T)~\lambda b_2\,((b_{1})~b_{2})~K}[M]$
      $=\underline{K}[(M)~\sigma(T)]$,
      which, by the induction hypothesis, \toblinred-reduces to
      $K[(M')~\sigma(T)]$
      $=\underline{\lambda b_1\,(T)~\lambda b_2\,((b_{1})~b_{2})~K}[M']$.
      \qedhere
  \end{itemize}
\end{proof}

\begin{lemma}\label{lem:continuation-linearity}
  The following relations hold.
  \begin{itemize}
    \item $\underline{K}[M_{1}+M_{2}]\stolinred\underline{K}[M_{1}]+\underline{K}[M_{2}]$
    \item $\underline{K}[\alpha.M]\stolinred\alpha.\underline{K}[M]$
    \item $\underline{K}[0]\stolinred0$
  \end{itemize}
\end{lemma}
\begin{proof}
  We prove each statement by induction on $K$, using Lemma \ref{lem:continuation-step}
  where necessary. We prove only the first statement, as the others
  are similar.
  \begin{itemize}
    \item Case $k$. Then $\underline{k}[M_{1}+M_{2}]=M_{1}+M_{2}=\underline{k}[M_{1}]+\underline{k}[M_{2}]$.
    \item Case $\lambda b\,((B)~b)~K$. Then 
      $\underline{\lambda b\,((B)~b)~K}[M_{1}+M_{2}]$
      $=\underline{K}[(\psi(B))~(M_{1}+M_{2})]$,
      which, by Lemma \ref{lem:continuation-step}, \tolinred-reduces to
      $\underline{K}[(\psi(B))~M_{1}+(\psi(B))~M_{2}]$,
      which, by the induction hypothesis, \stolinred-reduces to
      $\underline{K}[(\psi(B))~M_{1}]+\underline{K}[(\psi(B))~M_{2}]$
      $=\underline{\lambda b\,((B)~b)~K}[M_{1}]+\underline{\lambda b\,((B)~b)~K}[M_{2}]$.

    \item Case $\lambda b_1\,(S)~\lambda b_2\,((b_{1})~b_{2})~K$. Then
      the term
      $\underline{\lambda b_1\,(S)\lambda b_2\,((b_{1})~b_{2})~K)}[M_{1}+M_{2}]$
      is equal to
      $\underline{K}[(M_{1}+M_{2})~\sigma(S)]$
      which, by Lemma \ref{lem:continuation-step}, \tolinred-reduces to
      $\underline{K}[(M_{1})~\sigma(S)+(M_{2})~\sigma(S)]$,
      which, by the induction hypothesis, \stolinred-reduces to
      $\underline{K}[(M_{1})~\sigma(S)]+\underline{K}[(M_{2})~\sigma(S)]$
      which is equal to
      $\underline{\lambda b_1\,(S)~\lambda b_2\,((b_{1})~b_{2})~K}[M_{1}]+\underline{\lambda b_1\,(S)~\lambda b_2\,((b_{1})~b_{2})~K}[M_{2}]$.
      \qedhere
  \end{itemize}
\end{proof}

\begin{lemma}\label{lem:suspension-step}
  For any suspension $T$, if $T\toalgred T'$
  then $\sigma(T)\tolinred\sigma(T')$.
\end{lemma}
\begin{proof}
  By induction on the reduction rule. Since $T$ terms do not contain
  applications, the only cases possible are $L\cup\xi$, which are common
  to both languages.
  \begin{itemize}
    \item Case $L$. Using linearity of $\sigma$. We give the following example.
      $\sigma(T_{1}+(T_{2}+T_{3}))$
      $=\sigma(T_{1})+(\sigma(T_{2})+\sigma(T_{3}))$
      $\tolinred(\sigma(T_{1})+\sigma(T_{2}))+\sigma(T_{3})$
      $=\sigma((T_{1}+T_{2})+T_{3})$.

    \item Case $\xi$. Using linearity and the induction hypothesis. We give
      the following example. Consider the case $T_{1}+T_{2}\toalgred T_{1}'+T_{2}$
      with $T_{1}\toalgred T_{1}'$. Then 
      $\sigma(T_{1}+T_{2})$
      $\sigma(T_{1})+\sigma(T_{2})$,
      which, by the induction hypothesis, \tolinred-reduces to
      $\sigma(T_{1}')+\sigma(T_{2})$
      $=\sigma(T_{1}'+T_{2})$.
      \qedhere
  \end{itemize}
\end{proof}

We now have the tools to prove the Lemma~\ref{lem:inverse-step}.

\begin{proof}[\bf Proof of Lemma~\ref{lem:inverse-step}]
  By induction on the reduction rule, using Lemmas \ref{lem:substitution-lemma},
  \ref{lem:continuation-substitution}, \ref{lem:continuation-step},
  \ref{lem:continuation-linearity} and \ref{lem:suspension-step} where
  necessary.
  \begin{itemize}
    \item Case $\beta_{n}$. There are several sub-cases.
      \begin{itemize}
	\item Case $(\lambda b\,((B_{1})~b)~K)~B_{2}\tobn ((B_{1})~B_{2})~K$. Then 
	  $\overline{(\lambda b\,((B_{1})~b)~K)~B_{2}}$
	  is equal to
	  $\underline{\lambda b\,((B_{1})~b)~K}[\psi(B_{2})]$
	  $=\underline{K}[(\psi(B_{1}))~\psi(B_{2})]$
	  $=\overline{((B_{1})~B_{2})~K}$.

	\item Case $(\lambda b_1\,(S)~\lambda b_2\,((b_{1})~b_{2})~K)~B_{1}\tobn (S)~\lambda b_{2}\,((B_{1})~b_{2})~K$. Then we have that
	  $\overline{(\lambda b_1\,(S)~\lambda b_2\,((b_{1})~b_{2})~K)~B_{1}}$
	  $=\!\underline{\lambda b_1\,(S)\,\lambda b_2\,((b_{1})~b_{2})~K}[\psi(B_{1})]$
	  $=\!\underline{K}[(\psi(B_{1}))~\sigma(S)]$
	  $=\underline{\lambda b_2\,((B_{1})~b_{2})~K}[\sigma(S)]$
	  $=\overline{(S)~\lambda b_2\,((B_{1})~b_{2})~K}$.

	\item Case $(\lambda k\,C)~K\tobn C[k:=K]$. Then
	  $\overline{(\lambda k\,C)~K}$
	  $=\underline{K}[\overline{C}]$,
	  which, by Lemma \ref{lem:continuation-substitution}, is equal to
	  $\overline{C[k:=K]}$

	\item Case $((\lambda x\,S)~B)~K\tobn S[x:=B]K$. Then $\psi(B)$ is a
	  base term, and hence $(\lambda x\,\sigma(S))~\psi(B)\toblinred\sigma(S)[x:=\psi(B)]$,
	  so
	  $\overline{((\lambda x\,S)~B)~K}$
	  $=\underline{K}[(\lambda x\,\sigma(S))~\psi(B)]$,
	  which, by Lemma \ref{lem:continuation-step}, \toblinred-reduces to
	  $\underline{K}[\sigma(S)[x:=\psi(B)]]$,
	  which, by Lemma \ref{lem:substitution-lemma}, is equal to
	  $\underline{K}[\sigma(S[x:=B])]$
	  $=\overline{(S[x:=B])~K}$.
      \end{itemize}
    \item Case $A$. Since $B$ and $K$ are base terms, the only term that
      can match the rules is $(T)~K$. There are three sub-cases. 
      \begin{itemize}
	\item Case $(T_{1}+T_{2})~K\toalgred (T_{1})~K+(T_{2})~K$. Then 
	  $\overline{(T_{1}+T_{2})~K}$
	  $=\underline{K}[\sigma(T_{1})+\sigma(T_{2})]$,
	  which, by Lemma \ref{lem:continuation-linearity}, \stolinred-reduces to
	  $\underline{K}[\sigma(T_{1})]+\underline{K}[\sigma(T_{2})]$
	  $=\overline{(T_{1})~K+(T_{2})~K}$.
	\item Case $(\alpha.T)~K\toalgred\alpha.((T)~K)$. Then 
	  $\overline{(\alpha.T)~K}$,
	  $=\underline{K}[\alpha.\sigma(T)]$
	  which, by Lemma \ref{lem:continuation-linearity}, \stolinred-reduces to
	  $\alpha.\underline{K}[\sigma(T)]$
	  $=\overline{\alpha.((T)~K)}$.
	\item Case $(0)~K\toalgred0$. Then 
	  $\overline{(0)~K}$,
	  $=\underline{K}[0]$
	  which, by Lemma \ref{lem:continuation-linearity}, \stolinred-reduces to
	  $0$
	  $=\overline{0}$.
      \end{itemize}
    \item Case $L$. Since the rules in $L$ are common to both languages and
      the inverse translation $\overline{D}$ distributes linearly over
      the computations, the proof for these cases is straightforward. We
      give the following example. Consider $D_{1}+(D_{2}+D_{3})\toalgred(D_{1}+D_{2})+D_{3}$.
      Then
      $\overline{D_{1}+(D_{2}+D_{3})}$
      $=\overline{D_{1}}+(\overline{D_{2}}+\overline{D_{3}})$,
      and this \tolinred-reduces to
      $(\overline{D_{1}}+\overline{D_{2}})+\overline{D_{3}}$
      $=\overline{(D_{1}+D_{2})+D_{3}}$.

    \item Case $\xi$. There are 4 sub-cases.
      \begin{itemize}
	\item Case $(T)~K\toalgred (T')~K$ and $T\toalgred T'$. Then $\sigma(T)\tolinred\sigma(T')$
	  by Lemma \ref{lem:suspension-step}, therefore 
	  $\overline{(T)~K}$
	  $=\underline{K}[\sigma(T)]$,
	  by Lemma \ref{lem:continuation-step}, \tolinred-reduces to
	  $\underline{K}[\sigma(T')]$
	  $=\overline{(T')~K}$.

	\item The other three cases are similar to each other. We give the following
	  example. Consider $D_{1}+D_{2}\toalgred D_{1}'+D_{2}$ and $D_{1}\toalgred D_{1}'$.
	  Then by the induction hypothesis $\overline{D_{1}}\tolinred\overline{D_{1}'}$,
	  therefore 
	  $\overline{D_{1}+D_{2}}$
	  $=\overline{D_{1}}+\overline{D_{2}}$
	  $\tolinred\overline{D_{1}}'+\overline{D_{2}}$
	  $=\overline{D_{1}'+D_{2}}$.
	  \qedhere
      \end{itemize}
  \end{itemize}
\end{proof}

\subsection{Proof of Lemma~\ref{lem:substitution-cps}}\label{proof:substitution-cps}~

\recap{Lemma}{lem:substitution-cps}
$\cps{M[x:=N]}=\cps{M}[x:=\cps N]$.
\begin{proof}
  Structural induction on $M$.
  \begin{itemize}
    \item $M=x$. Then $\cps{x[x:=N]}=\cps{N}=x[x:=\cps{N}]=\cps{x}[x:=\cps{N}]$.
    \item $M=y$. Then $\cps{y[x:=N]}=y=\cps{y}[x:=\cps{N}]$.
    \item $M=0$. Analogous to previous case.
    \item $M=\lambda y\,M'$. Then
      $\cps{(\lambda y\,M')[x:=N]}$
      $=\cps{\lambda y\,(M'[x:=N])}$,
      which, by the induction hypothesis, is equal to
      $\cont{\lambda y\,\cps{M'[x:=N]}}$
      $=\cont{\lambda y\,\cps{M'}[x:=\cps{N}]}$
      $=(\cont{\lambda y\,\cps{M'}})[x:=\cps{N}]$
      $=\cps{M}[x:=\cps{N}]$.
    \item $M\,=\,(N_1)\ N_2$. Then 
      $\cps{M[x:=N]}$
      $\,=\,\cps{((N_1)~N_2)[x:=N]}$,
      which is equal to the\break term
      $\cps{(N_1[x:=N])~N_2[x:=N]}$,
      and this, by the induction hypothesis, is equal to
      $\lambda f\,(\cps{N_1[x:=N]})~\lambda g\,((g)~\cps{N_2[x:=N]})~f$ 
      $=\lambda f\,(\cps{N_1}[x:=\cps{N}])~\lambda g\,((g)$
      $\cps{N_2}[x:=\cps{N}])~f$
      $=(\lambda f\,(\cps{N_1})~\lambda g\,((g)~\cps{N_2})~f)[x:=\cps{N}]$
      $=\cps{(N_1)~N_2}[x:=\cps{N}]$
      $=\cps{M}[x:=\cps{N}]$.
    \item $M=\alpha.M'$. Then
      $\cps{M[x:=N]}$
      $=\cps{(\alpha.M')[x:=N]}$
      $=\cps{\alpha.(M'[x:=N])}$,
      which, by the induction hypothesis, is equal to
      $\lambda f\, (\alpha.\cps{M'[x:=N]))}) ~f$
      $=\lambda f\, (\alpha.\cps{M'}[x:=\cps{N}])~f$
      $=\lambda f\, (\alpha.\cps{M'})~f[x:=\cps{N}]$
      $=\cps{\alpha.M'}[x:=\cps{N}]$
      $=\cps{M}[x:=\cps{N}]$.
    \item $M=N_1+N_2$. Then 
      $\cps{M[x:=N]}$
      $= \cps{(N_1+N_2)[x:=N]}$
      which is equal to the term
      $\cps{N_1[x\!:=\!N]\!+\!N_2[x\!:=\!N]}$,
      which, by the induction hypothesis, is 
      $\lambda f\,(\cps{N_1[x\!:=N]}+\cps{N_2[x:=N]})~f$
      $=\lambda f\,
      (\cps{N_1}[x:=\cps{N}]+\cps{N_2}[x:=\cps{N}])~f$, which is
      $\lambda f\, ((\cps{N_1}+\cps{N_2})~f [x:=\cps{N}]$
      $=\cps{N_1+N_2}[x:=\cps{N}]$
      $=\cps{M}[x:=\cps{N}]$.
      \qedhere
  \end{itemize}
\end{proof}

\subsection{Proof of Lemma~\ref{lem:lemma2cps}}\label{proof:lemma2cps}~

\recap{Lemma}{lem:lemma2cps}
If $K$ is a base term,  for any term $M$
$(\cps{M})~K\stoblinred M:K$.
\begin{proof}
  Structural induction on $M$.
  \begin{itemize}
    \item $M=x$. Then $(\cps{x})~K=(x)~K = x:K$.
    \item $M=\lambda x\,N$. Then $(\cps{\lambda x\,N})~K=(\cont{\lambda x\,\cps{N}})~K$ and by definition of $\Phi$ this is equal to $(\cont{\Phi(\lambda x\,N)})~K \toblinred  (K) ~\Phi(\lambda x\,N)= \lambda x\,N : K$.
    \item $M=0$. Then $(\cps{0})~K = (\lambda f\, (0)~f)~K \toblinred (0)~K  \toblinred 0 = 0:K$.
    \item $M = M'+N$. Then $(\cps{M'+N})~K =(\lambda f\, (\cps{M'}+\cps{N})~f)~K\toblinred (\cps{M'}+\cps{N})~K $  which $\toblinred$-reduces by the induction hypothesis to $M':K+N:K = M'+N:K$.
    \item $M = \alpha.N$. Then $(\cps{\alpha.N})~K = (\lambda f\, (\alpha.\cps{N})~f)~K \toblinred  (\alpha. \cps{N})~K$ which $\toblinred$-reduces to $\alpha . (\cps{N})~K)$ and this, by the induction hypothesis, $\toblinred$-reduces to $ \alpha.(N:K) = \alpha.N : K$. 
    \item $M=(M')~N$. Then $(\cps{(M')~N})~K=(\lambda f\,(\cps{M'})~\lambda g\,((g)~\cps{N})~f)~K$ which $\toblinred$-reduces to $(\cps{M'})~\lambda g\,((g)~\cps{N})~K$. Note that $\lambda g\,((g)~\cps{N})~K$ is a value, so by the induction hypothesis the above term reduces to $M':\lambda g\,((g)~\cps{N})~K$. 
      We do a second induction, over $M'$, to prove that $M':\lambda g\,((g)~\cps{N})~K\stoblinred (M')~N:K$.
      \begin{itemize} 
	\item If $M'=(M_1)~M_2$, then $M':\lambda g\,((g)~\cps{N})~K=((M_1)~M_2)~N:K=(M')~N:K$.
	\item If $M'=x$ then $M':\lambda g\,((g)~\cps{N})~K=(x)~\lambda g\,((g)~\cps{N})~K=(M')~N:K$.
	\item If $M'=\lambda x\,M_1$ then $M':\lambda g\,((g)~\cps{N})~K=(\lambda g\,((g)~\cps{N})~K)~\Phi(M')$ which $\toblinred$-reduces to $((\Phi(M'))~\cps N)~K= (M')~N:K$.
	\item If $M' = \alpha.M_1$, then $\alpha.M_1:\lambda g\,((g)~\cps{N})~K=\alpha.(M_1:\lambda g\,((g)~\cps{N})~K)$  which $\stoblinred$-reduces by the induction hypothesis to $\alpha.((M_1)~N:K) = (\alpha.M_1)~N:K = (M')~N : K$.
	\item If $M' = M_1+M_2$, then $M':\lambda g\,((g)~\cps{N})~K=M_1+M_2 :\lambda g\,((g)~\cps{N})~K$ which is equal to $M_1 :\lambda g\,((g)~\cps{N})~K+M_2 :\lambda g\,((g)~\cps{N})~K$ which $\stoblinred$-reduces by the induction hypothesis to $(M_1)~N:K+(M_2)~N : K = (M_1+M_2)~N:K = (M')~N : K$.
	\item If $M'=0$ then $M':\lambda g\,((g)~\cps{N})~K= 0 :\lambda g\,((g)~\cps{N})~K = 0= (0)~N:K = (M')~N : K$.\qedhere
      \end{itemize}
  \end{itemize}
\end{proof}

\subsection{Proof of Lemma~\ref{lem:lemma3cps}}\label{proof:lemma3cps}~

\recap{Lemma}{lem:lemma3cps}
If $M\tobalgred N$ then for all $K$ base term, $M:K\stoblinred N:K$
\begin{proof}
  Case by case on the rules of $\lambda_{\textit{alg}}$.
  \begin{description}
    \item[Rule $\beta_v$]
      $(\lambda x\ M)\ N :K$
      $=((\Phi(\lambda x\ M))~\cps N)~K$
      $=((\lambda x\ (\cps M))~\cps N)~K$.
      Since $\cps N$ is a base term, this last term \toblinred-reduces to
      $\cps{M}[x:=(\cps N)]\,K$,
      which, by Lemma~\ref{lem:substitution-cps}, is equal to
      $\cps{M[x:=N]}\,K$,
      and this, by Lemma~\ref{lem:lemma2cps}, \stoblinred-reduces to
      $M[x:=N]:K$.
    \item[Rules $A$]~
      \begin{itemize}
	\item Let  $(M+N)~L \tobalgred (M) ~L+ (N)~L$. $(M+N)~L : K = ((M) ~L+ (N)~L):K $.
	\item Let  $(\alpha.M)~N \tobalgred \alpha.(M) ~N$. $(\alpha.M)~N :K= \alpha.(M) ~N:K$
	\item Let  $(0)~N \tobalgred 0$.  $(0)~N :K= 0 = 0:K$
      \end{itemize}
    \item[Rules $F$ and $S$]~
      \begin{itemize}
	\item $\alpha.(M+N)\tobalgred \alpha.M+\alpha.N$. Then $\alpha.(M+N):K=\alpha.(M:K+N:K)\toblinred\alpha.(M:K)+\alpha.(N:K)=\alpha.M+\alpha.N:K$.
	\item $\alpha.M+\beta.M\tobalgred(\alpha+\beta).M$. Then $\alpha.M+\beta.M:K=\alpha.(M:K)+\beta.(M:K)\toblinred(\alpha+\beta).(M:K)=(\alpha+\beta).M:K$.
	\item $\alpha.M+M\tobalgred(\alpha+1).M$. Then  $\alpha.M+M:K=\alpha.M:K+M:K=\alpha.(M:K)+M:K\toblinred(\alpha+1).(M:K)=(\alpha+1).M:K$.
	\item $M+M\tobalgred(1+1).M$. Then  $M+M:K=M:K+M:K\toblinred(1+1).(M:K)=(1+1).M:K$.
	\item $0+M\tobalgred M$. Then  $0+M:K=(0:K)+(M:K)=0+(M:K)\toblinred M:K$.
	\item $\alpha.(\beta.M)\tobalgred(\alpha\beta).M$. Then  $\alpha.(\beta.M):K=\alpha.(\beta.M:K)=\alpha.(\beta.(M:K))$ which $\toblinred$-reduces to $(\alpha\beta).(M:K)=(\alpha\beta).M:K$.
	\item $1.M\tobalgred M$. Then $1.M:K=1.(M:K)\toblinred M:K$.
	\item $0.M\tobalgred 0$. Then  $0.M:K=0.(M:K)\toblinred 0=0:K$.
	\item $\alpha.0\tobalgred 0$. Then  $\alpha.0:K=\alpha.(0:K)=\alpha.0\toblinred 0=0:K$.
      \end{itemize}
    \item[Rules $Asso$ and $Com$]~
      \begin{itemize}
	\item $M+(N+L)\tobalgred (M+N)+L$. Then $M+(N+L):K=M:K+(N+L:K)=M:K+(N:K+L:K)\toblinred (M:K+N:K)+L:K=M+N:K+L:K=(M+N)+L:K$.
	\item $M+N\tobalgred N+M$. Then $M+N:K=M:K+N:K\toblinred N:K+M:K=N+M:K$.
      \end{itemize}
    \item[Rules $\xi$] Assume $M\tobalgred M'$, and that for all $K$ base term, $M:K\stoblinred M':K$. We show that the result also holds for each contextual rule.
      \begin{itemize}
	\item $M+N\tobalgred M'+N$. Then $M+N:K=M:K+N:K\stoblinred M':K+N:K=M'+N:K$.
	\item $N+M\tobalgred N+M'$, analogous to previous case.
	\item $\alpha.M\tobalgred\alpha.M'$. Then $\alpha.M:K=\alpha.(M:K)\stoblinred \alpha.(M':K)=\alpha.M':K$.
	\item $(M)~N\tobalgred (M')~N$ Case by case:
	  \begin{itemize}
	    \item $M=B$. Absurd since a base term cannot reduce.
	    \item $M=\alpha.M_1$. Case by case on the possible $\tobalgred$-reductions of $M$:
	      \begin{itemize}
		\item $M'=\alpha.M_1'$ with $M_1\tobalgred M_1'$. Then $(\alpha.M_1)~N:K=\alpha.(M_1)~N:K$ $=\alpha.((M_1)~N:K)$ which, by the induction hypothesis, $\toblinred$-reduces to $\alpha.((M_1')~N:K)=\alpha.(M_1')~N:K=(\alpha.M_1')~N:K$.
		\item $M=\alpha.(\beta.M_3)$ and $M'=(\alpha\beta).M_3$. Then $(\alpha.(\beta.M_3))~N:K=\alpha.(\beta.((M_3)~N:K))\toblinred(\alpha\beta).((M_3)~N:K)=((\alpha\beta).M_3)~N:K$.
		\item $M=\alpha.(L_1+L_2)$ and $M'=\alpha.L_1+\alpha.L_2$. Then $(\alpha.(L_1+L2))~N:K=\alpha.((L_1)~N:K+(L_2)~N:K)\toblinred\alpha.((L_1)~N:K)+\alpha.((L_2)~N:K)=(\alpha.L_1+\alpha.L2)~N:K$.
		\item $\alpha=1$ and $M'=M_1$. Then $(1.M_1)~N:K=1.((M_1)~N:K)$ and this \toblinred-reduces to $(M_1)~N:K$.
		\item $\alpha=0$ and $M'=0$. Then $(0.M_1)~N:K=0.((M_1)~N:K)\toblinred 0=(0)~N:K$.
		\item $M_1=0$ and $M'=0$. Then $(\alpha.0)~N:K=\alpha.((0)~N:K)=\alpha.0\toblinred 0=(0)~N:K$.
	      \end{itemize}
	    \item $M=M_1+M_2$. Case by case on the possible $\tobalgred$-reductions of $M$:
	      \begin{itemize}
		\item $M'=M_1'+M_2$ with $M_1\tobalgred M'_1$. Then $(M_1+M_2)~N:K=(M_1)~N:K+(M_2)~N:K$ which by the induction hypothesis $\toblinred$-reduces to $(M_1')~N:K+(M_2)~N:K=(M_1'+M_2)~N:K$.
		\item $M'=M_1+M_2'$ with $M_2\tobalgred M'_2$. Analogous to previous case.
		\item $M_2=L_1+L_2$ and $M'=(M_1+L_1)+L_2$. Then $(M_1+(L_1+L_2))~N:K=(M_1)~N:K+((L_1)~N:K+(L_2)~N:K)$ which $\toblinred$-reduces to $((M_1)~N:K+(L_1)~N:K)+(L_2)~N:K=((M_1+L_1)+L_2)~N:K$.
		\item $M_1=L_1+L_2$ and $M'=L_1+(L_2+M_2)$. Analogous to previous case.
		\item $M'=M_2+M_1$. Then $(M_1+M_2)~N:K=(M_1)~N:K+(M_2)~N:K\toblinred (M_2)~N:K+(M_1)~N:K=(M_2+M_1)~N:K$.
		\item $M_1=\alpha.M_3$, $M_2=\beta.M_3$ and $M'=(\alpha+\beta).M_3$. Then $(\alpha.M_3+\beta.M_3)~N:K=\alpha.((M_3)~N:K)+\beta.((M_3)~N:K)\toblinred (\alpha+\beta).((M_3)~N:K)=((\alpha+\beta).M_3)~N:K$.
		\item $M_1=\alpha.M_3$, $M_2=M_3$ and $M'=(\alpha+1).M_3$. Analogous to previous case.
		\item $M_1=M_2$ and $M'=(1+1).M_1$. Analogous to previous case.
	      \end{itemize}
	    \item $M=0$. Absurd since $0$ does not reduce.
	    \item $M=(M_1)~M_2$. Then $((M_1)~M_2)~N:K=(M_1)~M_2:\lambda g\,((g)~\cps{N})~K$, which by the induction hypothesis $\toblinred$-reduces to $M':\lambda g\,((g)~\cps{N})~K$.
	      We do a second induction, over $M'$, to prove that $M':\lambda g\,((g)~\cps{N})~K$ \stoblinred-reduces to $(M')~N:K$.
	      \begin{itemize} 
		\item If $M'=(M'_1)~M'_2$, then $M':\lambda g\,((g)~\cps{N})~K=((M'_1)~M'_2)~N:K=(M')~N:K$.
		\item If $M'=x$ then $M':\lambda g\,((g)~\cps{N})~K=(x)~\lambda g\,((g)~\cps{N})~K=(M')~N:K$.
		\item If $M'=\lambda x\,M'_1$ then $M':\lambda g\,((g)~\cps{N})~K$ is \
		  $(\lambda g\,((g)~\cps{N})~K)~\Phi(M')\toblinred((\Phi(M'))~\cps N)~K= (M')~N:K$.
		\item If $M' = \alpha.M'_1$, then $\alpha.M'_1:\lambda g\,((g)~\cps{N})~K=\alpha.(M'_1:\lambda g\,((g)~\cps{N})~K)$  which $\stoblinred$-reduces by the induction hypothesis to $\alpha.((M'_1)~N:K) = (\alpha.M'_1)~N:K = (M')~N : K$.
		\item If $M' = M'_1+M'_2$, then $M':\lambda g\,((g)~\cps{N})~K=M'_1+M'_2 :\lambda g\,((g)~\cps{N})~K$ which is equal to $M'_1 :\lambda g\,((g)~\cps{N})~K+M'_2 :\lambda g\,((g)~\cps{N})~K$ which $\stoblinred$-reduces by the induction hypothesis to $(M'_1)~N:K+(M'_2)~N : K = (M'_1+M'_2)~N:K = (M')~N : K$.
		\item If $M'=0$ then $M':\lambda g\,((g)~\cps{N})~K= 0 :\lambda g\,((g)~\cps{N})~K = 0= (0)~N:K = (M')~N : K$.\qedhere
	      \end{itemize}
	  \end{itemize}
      \end{itemize}
  \end{description} 
\end{proof}

\subsection{Proof of Lemma \ref{lem:inverse-term-a}}\label{proof:inverse-term-a}~
\mynobreakpar
\recap{Lemma}{lem:inverse-term-a} For any term $M$, $\overline{(\cps M)~k}=M$.\mynobreakpar
\begin{proof}
  By induction on $M$.\mynobreakpar
  \begin{itemize}
    \item Case $x$. Then $\sigma(\cps x)=\sigma(x)=x$.
    \item Case $\lambda x\,M$. Then 
      $\sigma(\cps{\lambda x\,M})$
      $=\sigma(\lambda k\,(k)~\lambda x\,\cps M)$
      $=\overline{(k)~\lambda x\,\cps M}$
      $=\underline{k}[\phi(\lambda x\,\cps M)]$
      $=\phi(\lambda x\,\cps M)$
      $=\lambda x\,\sigma(\cps M)$,
      which, by the induction hypothesis, is equal to
      $\lambda x\,M$.

    \item Case $(M)~N$. Then 
      $\sigma(\cps{(M)~N})$
      $=\sigma(\lambda k\,(\cps M)~\lambda b.((b)~\cps N)~k)$
      which is equal to
      $\overline{(\cps M)~\lambda b.((b)~\cps N)~k}$
      $\,=\,\underline{\lambda b.((b)~\cps N)~k}[\sigma(\cps M)]$
      $\,=\,\underline{k}[(\sigma(\cps M))~\sigma(\cps N)]$, equal to
      $(\sigma(\cps M))~\sigma(\cps N)$,
      which, by the induction hypothesis, is equal to
      $(M)~N$.

    \item Case $0$. Then $\sigma(\cps 0)=\sigma(0)=0$
    \item Case $\alpha.M$. Then 
      $\sigma(\cps{\alpha.M})$
      $=\sigma(\lambda k\,(\alpha.\cps M)~k)$, which is equal to
      the term
      $\overline{(\alpha.\cps M)k}$
      $=\underline{k}[\sigma(\alpha.\cps M)]$
      $=\sigma(\alpha.\cps M)$
      $=\alpha.\sigma(\cps M)$,
      which, by the induction hypothesis, is equal to
      $\alpha.M$.

    \item Case $M+N$. Then
      $\sigma(\cps{M+N})$
      $=\sigma(\lambda k\,(\cps M+\cps N)~k)$
      $=\overline{(\cps M+\cps N)~k}$
      $=\underline{k}[\sigma(\cps M+\cps N)]$
      $=\sigma(\cps M+\cps N)$
      $=\sigma(\cps M)+\sigma(\cps N)$,
      which, by the induction hypothesis, is equal to
      $M+N$.
      \qedhere
  \end{itemize}
\end{proof}

\subsection{Proof of Lemma \ref{lem:inverse-value-a}}\label{proof:inverse-value-a}~

\recap{Lemma}{lem:inverse-value-a} For any value $V$, $\overline{V:k}=V$.
\begin{proof}
  By induction on $V$.
  \begin{itemize}
    \item Case $0$. Then $\overline{0:k}=\overline{0}=0$.
    \item Case $x$. Then $\overline{x:k}=\overline{(x)~k}=\underline{k}[x]=x$.
    \item Case $\lambda x\,M$. Then $\overline{\lambda x\,M:k}=\overline{(k)~\Phi(\lambda x\,M)}=\sigma(\lambda k\,(k)~\Phi(\lambda x\,M))=\sigma(\cps{\lambda x\,M})=\lambda x\,M$
      by Lemma \ref{lem:inverse-term-a}.
    \item Case $\alpha.V$. Then $\overline{\alpha.V:k}=\overline{\alpha.(V:k)}=\alpha.\overline{V:k}=\alpha.V$
      by the induction hypothesis.
    \item Case $U+V$. Then $\overline{U+V:k}=\overline{U:k+V:k}=\overline{U:k}+\overline{V:k}=U+V$
      by the induction hypothesis.
      \qedhere
  \end{itemize}
\end{proof}

\subsection{Proof of Lemma \ref{lem:inverse-step-a}}\label{proof:inverse-step-a}~

\recap{Lemma}{lem:inverse-step-a} For any computation $D$, if $D\toblinred D'$
then $\overline{D}\stobalgred\overline{D'}$. Also, if $D\tolinred D'$
then $\overline{D}\stoalgred\overline{D'}$.

\bigskip
In order to prove this lemma, we need intermediary results similar to Lemmas \ref{lem:substitution-lemma}, \ref{lem:continuation-composition}, \ref{lem:continuation-substitution}, \ref{lem:continuation-step}, \ref{lem:continuation-linearity}, and \ref{lem:suspension-step}.

\begin{lemma}
  \label{lem:substitution-lemma-a} The following equalities hold.
  \begin{myenumerate}
    \item {$\phi(B)[x:=\sigma(S)]=\phi(B[x:=S])$}
    \item $\sigma(T)[x:=\sigma(S)]=\sigma(T[x:=S])$
    \item {$\overline{C}[x:=\sigma(S)]=\overline{C[x:=S]}$}
    \item $\underline{K}[M][x:=\sigma(S)]=\underline{K[x:=S]}[M[x:=\sigma(S)]]$
  \end{myenumerate}
\end{lemma}
\begin{proof}
  We prove simultaneously the four properties by induction on the structure of $B$, $T$, $C$ and $K$.
  \begin{myenumerate}
    \item Cases for $B$.
      \begin{itemize}
	\item Case $\lambda y\,S$. Then 
	  $\phi(\lambda x\,S_{1})[x:=\sigma(S)]$
	  $=(\lambda y\,\sigma(S_{1}))[x:=\sigma(S)]$,
	  which, by the induction hypothesis, is equal to
	  $\lambda y\,\sigma(S_{1}[x:=S])$
	  $=\phi((\lambda y\,S_{1})[x:=S])$.
      \end{itemize}
    \item Cases for $T$.
      \begin{itemize}
	\item Case $x$. Then $\sigma(x)[x:=\sigma(S)]=x[x:=\sigma(S)]=\sigma(S)=\sigma(x[x:=S])$.
	\item Case $y\neq x$. Then $\sigma(y)[x:=\sigma(S)]=y[x:=\sigma(S)]=y=\sigma(y)=\sigma(y[x:=S])$.
	\item Case $\lambda k\,C$. Then 
	  $\sigma(\lambda k\,C)[x:=\sigma(S)]$
	  $=\overline{C}[x:=\sigma(S)]$,
	  which, by the induction hypothesis, is equal to
	  $\overline{C[x:=S]}$
	  $=\sigma((\lambda k\,C)[x:=S])$.

	\item Case $0$. Then $\sigma(0)[x:=\sigma(S)]=0=\sigma(0[x:=S])$.
	\item Case $\alpha.T$. Then 
	  $\sigma(\alpha.T)[x:=\sigma(S)]$
	  $=(\alpha.\sigma(T))[x:=\sigma(S)]$,
	  which, by the induction hypothesis, is equal to
	  $\alpha.\sigma(T[x:=S])$
	  $=\sigma((\alpha.T)[x:=S])$.

	\item Case $T_{1}+T_{2}$. Then 
	  $\sigma(T_{1}+T_{2})[x:=\sigma(S)]$
	  $=(\sigma(T_{1})+\sigma(T_{2}))[x:=\sigma(S)]$,
	  which, by the induction hypothesis, is equal to
	  $\sigma(T_{1}[x:=S])+\sigma(T_{2}[x:=S])$
	  $=\sigma((T_{1}+T_{2})[x:=S])$.
      \end{itemize}
    \item Cases for $C$.
      \begin{itemize}
	\item Case $(K)~B$. Then 
	  $\overline{(K)~B}[x:=\sigma(S)]$
	  $=\underline{K}[\phi(B)][x:=\sigma(S)]$,
	  which, by the induction hypothesis, is equal to
	  $\underline{K[x:=S]}[\phi(B)[x:=\sigma(S)]]$,
	  which, by the induction hypothesis, is equal to
	  $\underline{K[x:=S]}[\phi(B[x:=S])]$
	  $=\overline{((K)~B)[x:=S]}$.

	\item Case $((B)~S_{2})~K$. Then
	  $\overline{(B)~S_{2})~K}[x:=\sigma(S)]$
	  $=\underline{K}[(\psi(B))~\sigma(S_{2})][x:=\sigma(S)]$,
	  which, by the induction hypothesis, is equal to
	  $\underline{K[x:=S]}[((\psi(B))~\sigma(S_{2}))[x:=\sigma(S)]]$,
	  which, by the induction hypothesis, is equal to
	  $\underline{K[x:=S]}[(\psi(B[x:=S]))~\sigma(S_{2}[x:=S])]$
	  $=\overline{(((B)~S_{2})~K)[x:=S]}$.

	\item Case $(T)~K$. Then 
	  $\overline{(T)~K}[x:=\sigma(S)]$
	  $=\underline{K}[\sigma(T)][x:=\sigma(S)]$,
	  which, by the induction hypothesis, is equal to
	  $\underline{K[x:=S]}[\sigma(T)[x:=\sigma(S)]]$,
	  which, by the induction hypothesis, is equal to
	  $\underline{K[x:=S]}[\sigma(T[x:=S])]$
	  $=\overline{((T)~K)[x:=S]}$
      \end{itemize}
    \item Cases for $K$.
      \begin{itemize}
	\item Case $k$. Then $\underline{k}[M][x:=\sigma(S)]=M[x:=\sigma(S)]=\underline{k[x:=S]}[M[x:=\sigma(S)]]$.
	\item Case $\lambda b\,((b)~S_{2})~K$. Then 
	  $\underline{\lambda b\,((b)~S_{2})~K}[M][x:=\sigma(S)]$
	  $=\underline{K}[M\sigma(S_{2})][x:=\sigma(S)]$,
	  which, by the induction hypothesis, is 
	  $\underline{K[x:=S]}[((M)~\sigma(S_{2}))[x:=\sigma(S)]]$,
	  which, by the induction hypothesis, is equal to
	  $\underline{K[x:=S]}[M[x:=\sigma(S)]\sigma(S_{2}[x:=S])]$
	  $=\underline{(\lambda b\,((b)~S_{2})~K)[x:=S]}[M[x:=\sigma(S)]]$.
	  \qedhere
      \end{itemize}
  \end{myenumerate}
\end{proof}

\begin{lemma}
  \label{lem:continuation-composition-a} For all terms $M$ and continuations
  $K_{1}$ and $K_{2}$, we have,
  $\underline{K_{1}}[\underline{K_{2}}[M]]=\underline{K_{2}[k:=K_{1}]}[M]$.
\end{lemma}
\begin{proof}
  By induction on $K_{2}$.
  \begin{itemize}
    \item Case $k$. Then $\underline{K_{1}}[\underline{k}[M]]=\underline{K_{1}}[M]=\underline{k[k:=K_{1}]}[M]$.
    \item Case $\lambda b\,((b)~S)~K$. Then 
      $\underline{K_{1}}[\underline{\lambda b\,((b)~S)~K}[M]]$
      $=\underline{K_{1}}[\underline{K}[(M)~\sigma(S)]]$,
      which, by the induction hypothesis, is equal to
      $\underline{K[k:=K_{1}]}[(M)~\sigma(S)]$
      $=\underline{(\lambda b\,((b)~S)~K)[k:=K_{1}]}[M]$.
      \qedhere
  \end{itemize}
\end{proof}

\begin{lemma}
  {\label{lem:continuation-substitution-a} For all $K$ and
  $C$, $\underline{K}[\overline{C}]=\overline{C[k:=K]}$}.
\end{lemma}
\begin{proof}
  By induction on the structure of $C$, using Lemma \ref{lem:continuation-composition-a}
  where necessary.
  \begin{itemize}
    \item Case $(K_{2})~B$. Then 
      $\underline{K}[\overline{(K_{2})~B}]$
      $=\underline{K}[\underline{K_{2}}[\phi(B)]]$,
      which, by Lemma \ref{lem:continuation-composition-a}, is equal to
      $\underline{K_{2}[k:=K]}[\phi(B)]$
      $=\overline{((K_{2})~B)[k:=K]}$.

    \item Case $((B)~S)~K_{2}$. Then 
      $\underline{K}[\overline{((B)~S)~K_{2}}]$
      $=\underline{K}[\underline{K_{2}}[(\phi(B))~\sigma(S)]]$,
      which, by Lemma \ref{lem:continuation-composition-a}, is equal to
      $\underline{K_{2}[k:=K]}[(\phi(B))~\sigma(S)]$
      $=\overline{(((B)~S)~K_{2})[k:=K]}$.

    \item Case $(T)~K_{2}$. Then 
      $\underline{K}[\overline{(T)~K_{2}}]$
      $=\underline{K}[\underline{K_{2}}[\sigma(T)]]$,
      which, by Lemma \ref{lem:continuation-composition-a}, is equal to
      $\underline{K_{2}[k:=K]}[\sigma(T)]$
      $=\overline{((T)~K_{2})[k:=K]}$.
      \qedhere
  \end{itemize}
\end{proof}

\begin{lemma}
  \label{lem:continuation-step-a}For any continuation $K$ and term
  $M$, if $M\tobalgred M'$ then $\underline{K}[M]\tobalgred\underline{K}[M']$.
\end{lemma}
\begin{proof}
  By induction on the structure of $K$.
  \begin{itemize}
    \item Case $k$. Then $\underline{k}[M]=M\tobalgred M'=\underline{k}[M']$.
    \item Case $\lambda b\,((b)~S)~K$. Then we have $M\sigma(S)\tobalgred M'\sigma(S)$, and 
      $\underline{\lambda b\,((b)~S)~K}[M]$
      $=\underline{K}[(M)~\sigma(S)]$,
      and this, by the induction hypothesis, \tobalgred-reduces to 
      $\underline{K}[(M')~\sigma(S)]$
      $=\underline{\lambda b\,((b)~S)~K}[M']$.
      \qedhere
  \end{itemize}
\end{proof}

\begin{lemma}
  \label{lem:continuation-linearity-a}
  For any continuation $K$, scalar $\alpha$ and terms $M$, $M_1$ and $M_2$, the following relations hold.
  \begin{itemize}
    \item $\underline{K}[M_{1}+M_{2}]\stoalgred\underline{K}[M_{1}]+\underline{K}[M_{2}]$
    \item $\underline{K}[\alpha.M]\stoalgred\alpha.\underline{K}[M]$
    \item $\underline{K}[0]\stoalgred0$
  \end{itemize}
\end{lemma}

\begin{proof}
  We prove each statement by induction on $K$, using Lemma \ref{lem:continuation-step-a}
  where necessary.We prove only the first statement, as the others
  are similar.
  \begin{itemize}
    \item Case $k$. Then $\underline{k}[M_{1}+M_{2}]=M_{1}+M_{2}=\underline{k}[M_{1}]+\underline{k}[M_{2}]$.
    \item Case $\lambda b\,((b)~S)~K$. Then 
      $\underline{\lambda b\,((b)~S)~K}[M_{1}+M_{2}]$
      $=\underline{K}[(M_{1}+M_{2})~\sigma(S)]$,
      which, by Lemma \ref{lem:continuation-step}, \toalgred-reduces to
      $\underline{K}[(M_{1})~\sigma(S)+(M_{2})~\sigma(S)]$,
      and this, by the induction hypothesis, \stoalgred-reduces to
      $\underline{K}[(M_{1})~\sigma(S)]+\underline{K}[(M_{2})~\sigma(S)]$
      $=\underline{\lambda b\,((b)~S)~K}[M_{1}]+\underline{\lambda b\,((b)~S)~K}[M_{2}]$
      \qedhere
  \end{itemize}
\end{proof}

\begin{lemma}
  \label{lem:suspension-step-a} For any suspension $T$, if $T\tolinred T'$
  then $\sigma(T)\toalgred\sigma(T')$.
\end{lemma}
\begin{proof}
  By induction on the reduction rule. Since $T$ terms do not contain
  applications, the only cases possible are $L\cup\xi$, which are common
  to both languages.
  \begin{itemize}
    \item Case $L$. Using linearity of $\sigma$. We give the following example.
      $\sigma(T_{1}+(T_{2}+T_{3}))$
      $=\sigma(T_{1})+(\sigma(T_{2})+\sigma(T_{3}))$
      $\toalgred(\sigma(T_{1})+\sigma(T_{2}))+\sigma(T_{3})$
      $=\sigma((T_{1}+T_{2})+T_{3})$.

    \item Case $\xi$. Using linearity and the induction hypothesis. We give
      the following example. Consider the case $T_{1}+T_{2}\tolinred T_{1}'+T_{2}$
      with $T_{1}\tolinred T_{1}'$. Then 
      $\sigma(T_{1}+T_{2})$
      $=\sigma(T_{1})+\sigma(T_{2})$,
      which, by the induction hypothesis, \toalgred-reduces to
      $\sigma(T_{1}')+\sigma(T_{2})$
      $=\sigma(T_{1}'+T_{2})$.
      \qedhere
  \end{itemize}
\end{proof}

We now have the tools to prove the Lemma~\ref{lem:inverse-step-a}.

\begin{proof}[\bf Proof of Lemma~\ref{lem:inverse-step-a}]
  By induction on the reduction rule, using Lemmas \ref{lem:substitution-lemma-a},
  \ref{lem:continuation-substitution-a}, \ref{lem:continuation-step-a},
  \ref{lem:continuation-linearity-a} and \ref{lem:suspension-step-a}
  where necessary. The rules $\xi_{\lambda_{lin}}$ and
  $A_{r}$ are not applicable since arguments in the target language
  are always base terms.
  \begin{itemize}
    \item Case $\beta_{v}$. There are several sub-cases.
      \begin{itemize}
	\item Case $(\lambda b\,((b)~S)~K)~B\tobv ((B)~S)~K$. Then 
	  $\overline{(\lambda b\,((b)~S)~K)~B}$
	  which is equal to
	  $\underline{\lambda b\,((b)~S)~K}[\phi(B)]$
	  $=\underline{K}[(\phi(B))~\sigma(S)]$
	  $=\overline{((B)~S)~K}$.

	\item Case $(\lambda k\,C)~K\tobv C[k:=K]$. Then
	  $\overline{(\lambda k\,C)~K}$
	  $=\underline{K}[\overline{C}]$,
	  which, by Lemma \ref{lem:continuation-substitution-a}, is equal to
	  $\overline{C[k:=K]}$.

	\item Case $((\lambda x\,S)~S_{2})~K\tobv (S[x:=S_{2}])~K$. Then $(\lambda x\,\sigma(S))~\sigma(S_{2})\toalgred\sigma(S)[x:=\sigma(S_{2})]$,
	  and
	  $\overline{((\lambda x\,S)~S_{2})~K}$
	  $=\underline{K}[(\lambda x\,\sigma(S))~\sigma(S_{2})]$,
	  which, by Lemma \ref{lem:continuation-step-a}, \tolinred-reduces to
	  $\underline{K}[\sigma(S)[x:=\sigma(S_{2})]]$,
	  which, by Lemma \ref{lem:substitution-lemma-a}, is equal to
	  $\underline{K}[\sigma(S[x:=S_{2}])]$
	  $=\overline{(S[x:=S_{2}])~K}$.
      \end{itemize}
    \item Case $A_{l}$. Since $B$ and $K$ are base terms, the only term that
      can match the rules is $(T)~K$. There are three sub-cases. 
      \begin{itemize}
	\item Case $(T_{1}+T_{2})~K\tolinred (T_{1})~K+(T_{2})~K$. Then 
	  $\overline{(T_{1}+T_{2})~K}$
	  $=\underline{K}[\sigma(T_{1})+\sigma(T_{2})]$,
	  which, by Lemma \ref{lem:continuation-linearity-a}, \stoalgred-reduces to
	  $\underline{K}[\sigma(T_{1})]+\underline{K}[\sigma(T_{2})]$
	  $=\overline{(T_{1})~K+(T_{2})~K}$.
	\item Case $(\alpha.T)~K\tolinred\alpha.((T)~K)$. Then 
	  $\overline{(\alpha.T)~K}$,
	  $=\underline{K}[\alpha.\sigma(T)]$,
	  which, by Lemma \ref{lem:continuation-linearity-a}, \stoalgred-reduces to
	  $\alpha.\underline{K}[\sigma(T)]$
	  $=\overline{\alpha.((T)~K)}$.
	\item Case $(0)~K\tolinred0$. Then 
	  $\overline{(0)~K}$
	  $=\underline{K}[0]$,
	  which, by Lemma \ref{lem:continuation-linearity-a}, \stoalgred-reduces to
	  $0$
	  $=\overline{0}$.
      \end{itemize}
    \item Case $L$. Since the rules in $L$ are common to both languages and
      the inverse translation $\overline{D}$ distributes linearly over
      the computations, the proof for these cases is straightforward. We
      give the following example. Consider $D_{1}+(D_{2}+D_{3})\tolinred(D_{1}+D_{2})+D_{3}$.
      Then
      $\overline{D_{1}+(D_{2}+D_{3})}$
      $=\overline{D_{1}}+(\overline{D_{2}}+\overline{D_{3}})$
      $\toalgred(\overline{D_{1}}+\overline{D_{2}})+\overline{D_{3}}$,
      which is equal to
      $\overline{(D_{1}+D_{2})+D_{3}}$.

    \item Case $\xi$. There are 4 sub-cases.
      \begin{itemize}
	\item Case $(T)~K\tolinred (T')~K$ and $T\tolinred T'$. Then $\sigma(T)\toalgred\sigma(T')$
	  by Lemma \ref{lem:suspension-step-a}, therefore 
	  $\overline{(T)~K}$
	  $=\underline{K}[\sigma(T)]$,
	  which, by Lemma \ref{lem:continuation-step-a}, \toalgred-reduces to
	  $\underline{K}[\sigma(T')]$
	  $=\overline{(T')~K}$.

	\item The other three cases are similar to each other. We give the following
	  example. Consider $D_{1}+D_{2}\tolinred D_{1}'+D_{2}$ and $D_{1}\tolinred D_{1}'$.
	  Then by the induction hypothesis $\overline{D_{1}}\toalgred\overline{D_{1}'}$,
	  therefore 
	  $\overline{D_{1}+D_{2}}$
	  $=\overline{D_{1}}+\overline{D_{2}}$
	  $\toalgred\overline{D_{1}}'+\overline{D_{2}}$
	  $=\overline{D_{1}'+D_{2}}$.
	  \qedhere
      \end{itemize}
  \end{itemize}
\end{proof}

\end{document}